\documentclass[fleqn,usenatbib]{mnras}

\usepackage{newtxtext,newtxmath}

\usepackage[T1]{fontenc}
\usepackage{pdfpages}

\usepackage{graphicx}	
\usepackage{amsmath}	

\DeclareRobustCommand{\VAN}[3]{#2}
\let\VANthebibliography\thebibliography
\def\thebibliography{\DeclareRobustCommand{\VAN}[3]{##3}\VANthebibliography}

\newcommand{\eps}{erg s$^{-1}$}

\newcommand{\source}{Swift J1728.9--3613}

\title[\source]{Accretion Properties and Estimation of Spin of Galactic Black Hole Candidate \source ~with {\it NuSTAR} during its 2019 outburst}

\author[Heiland et al.]{
Skye R. Heiland,$^{1}$\thanks{E-mail: heilands@myumanitoba.ca}
Arka Chatterjee,$^{1}$\thanks{E-mail: arka.chatterjee@umanitoba.ca}
Samar Safi-Harb,$^{1}$
Arghajit Jana,$^{2}$
and Jeremy Heyl$^{3}$\\
$^{1}$Department of Physics and Astronomy, University of Manitoba, Winnipeg, Manitoba, R3T 2N2, Canada\\
$^{2}$Institute of Astronomy, National Tsing Hua University, Hsinchu, 300044, Taiwan\\
$^{3}$Department of Physics and Astronomy, University of British Columbia, Vancouver, V6T 1Z1, Canada
}

\date{Accepted XXX. Received YYY; in original form ZZZ}

\pubyear{2023}

\begin{document}
\label{firstpage}
\pagerange{\pageref{firstpage}--\pageref{lastpage}}
\maketitle

\begin{abstract}
Black hole X-ray binaries (BHXRBs) play a crucial role in understanding the accretion of matter onto a black hole. Here, we focus on exploring the transient BHXRB \source~discovered by Swift/BAT and MAXI/GSC during its January 2019 outburst. We present measurements on its accretion properties, long time-scale variability, and spin. To probe these properties we make use of several NICER  observations and an unexplored data set from NuSTAR, as well as long term light curves from MAXI/GSC. In our timing analysis we provide estimates of the cross-correlation functions between light curves in various energy bands. In our spectral analysis we employ numerous phenomenological models to constrain the parameters of the system, including flavours of the relativistic reflection model {\sc Relxill} to model the Fe K$\alpha$ line and the $>15$ keV reflection hump. Our analysis reveals that: (i) Over the course of the outburst the total energy released was $\sim 5.2 \times 10^{44}$~ergs, corresponding to roughly 90\% the mass of Mars being devoured. (ii) We find a continuum lag of {\bf $8.4 \pm 1.9$} days between light curves in the $2-4$ keV and $10-20$ keV bands which could be related to the viscous inflow time-scale of matter in the standard disc. (iii) Spectral analysis reveals a spin parameter of $\sim 0.6 - 0.7$ with an inclination angle of $\sim 45^{\circ}-70^{\circ}$, and an accretion rate during the NuSTAR observation of $\sim 17\% ~L_{\rm Edd}$.
\end{abstract}




\section{Introduction}
Black hole X-ray binaries (BHXRBs) are a class of astrophysical objects in which a black hole and a main sequence star are gravitationally bound. These systems make ideal laboratories for studying the physics of accretion and matter in the presence of an extreme gravitational well, in addition to testing the predictions of general relativity. BHXRBs are further categorized into transient and persistent classes. The transient class remains dormant for most of its lifetime, except for occasional outbursts where the luminosities ($L_{\rm X}$) reach beyond $10^{35}$ \eps \citep[e.g.,][]{Tetarenko2016}. These outbursts and their spectral evolution are traced through the so-called `q' or hardness--intensity diagram \citep[e.g.,][]{Homan2001,RM06}. A full outburst starts with a low/hard state followed by an intermediate state reaching to the high/soft state while rising to the peak luminosity. During the declining phase, the path reverses ending with the hard state. `Failed' or hard/low state only outbursts are those where high/soft state remains absent \citep[see ][and references therein]{Alberta2021}. Energy dependent variations in count rates are a universal feature of outbursting BHXRBs, which are usually attributed to their accretion properties. A few major physical drivers of accretion are viscosity \citep[e.g.,][]{SS73,Smith2002_accretion,Chatterjee2020}, electron cloud temperature and optical depth \citep{ST80,ST85}, the Compton cooling rate of the cloud \citep[e.g.,][]{CT95,Ghosh2012,Garain2014}. In general, measurements of such parameters are performed using long term light-curve variations and detailed spectral studies using various phenomenological or physical models. 

Apart from accretion properties, outbursting BHXRBs also provide a chance to measure the intrinsic properties of the black hole itself, e.g. mass ($M_{\rm BH}$) and spin ($a^*$). The dynamical method, wherein mass is obtained using radial velocity measured from absorption lines in the spectrum from the companion star, provides the most accurate measure of the mass \citep{CJ2014}. One can also use dips and eclipses in the light curve to predict the mass ratio and inclination of the binary system \citep{Horne1985}. However, only around 20 BHXRBs are measured this way \citep[e.g.,][]{Corral2016}. Apart from the dynamical method, one can measure the mass using spectral fitting \citep[e.g.,][]{Kreidberg2012,Torres2019,Kubota1998,Shaposhnikov2007,Jana2022}  or the X-ray reverberation method \citep{Mastroserio2019}. Direct measurement of the spin could be carried out by imaging the shape of the photon sphere or radius of the innermost stable circular orbit (ISCO) \citep{Chan2013}. Using radio interferometry, the EHT-collaboration directly captured images of the supermassive black holes M 87* and our own Sgr A* (see \cite{EHT2019,EHT2022} and references therein). For galactic BHXRBs, spin is usually measured by X-ray spectroscopy (e.g \cite{Draghis2022}), though in principle X-ray interferometry could be used to produce EHT-style observations that directly image the shape of the BH shadow in some Seyfert galaxies \citep{Uttley2020,Hartog2020}.

Within X-ray spectroscopy there are two primary methods that have proved successful in measuring black hole spin, namely continuum fitting \citep[e.g.,][]{Zhang1997,McClintock2006,Steiner2014} and relativistic reflection \citep[e.g.,][]{Fabian1989,Miller2012,Reynolds2021}. In continuum fitting, the inner edge of the accretion disc or ISCO ($r_{\rm in}$) is measured 
through the inner disc temperature. The optically thick and geometrically thin accretion disc considers the relativistic form prescribed by \citep{Novikov1973} where the spin influences $r_{\rm in}$. For extreme prograde motion (i.e. $a\rightarrow 1$), $r_{\rm in}$, marginally stable and bound orbits and, the photon sphere merge into the surface of the event horizon. As the inner radius becomes smaller, the excess binding energy loss, spent into heating the accretion disc, becomes increasingly larger while approaching $r_{\rm in}$. The corresponding temperature $T_{\rm in}$ can then be measured using relativistic models like {\sc Kerrbb} \citep{Zhang1997,Li2005}. This method was employed to measure the spins of LMC~X--1 \citep{Gou2009,Mudambi2020}, H 1743-322 \citep{Steiner2009}, GRS 1915+105 \citep{Sree2020}, LMC~X--1 \citep{AJ2021d}, LMC~X--3 \citep{Bhuvana2021}, and MAXI~J1820+070 \citep{Zhao2021}.  

In contrast to continuum fitting, the relativistic reflection method instead models the spectral shape using the blurred iron K$\alpha$ line at 6.4~keV and associated reflection hump above 15~keV \citep{Fabian1989,George1991}. This fluorescent line distorts due to the gravitational redshift experienced by the reflected spectra of a spinning black hole, with maximal distortion occurring for a maximally spinning prograde orbit. The `reflection hump' arises due to the hard X-ray reflection from the accretion disc. Often the irradiation of the disc shows a profile steeper than $r^{-5}$, invoking a compact corona along the rotation axis of the black hole. This is the so-called `lamp-post' model of the coronal geometry \citep{MF2004}, and the height and compactness of this corona change alongside other spectral properties \citep{Fabian2015}. As the Doppler effect significantly modifies the observed spectrum with changing inclination \citep{Luminet1979,Viergutz1993}, these models calculate the parameters of the system and modify the iron line profile accordingly. A central assumption invoked in this method is that the inner radius of the accretion disk extends all the way to the inner most stable circular orbit or ISCO. This need not be the case, especially in the low/hard state \citep[e.g.][]{Zdziarski2020, Done2007}. The spectroscopic reflection model {\sc Relxill} \citep{Garcia2013, Dauser2014, Garcia2014} provides a measure of the iron abundance ($A_{\rm Fe}$) within the disc, and various flavours of disc/emissivity profiles can be assumed. Within the past few years, {\sc Relxill} has successfully measured the spin of many galactic black holes, such as MAXI~J1535--571 \citep{Miller2018}, Cygnus X--1 \citep{Tomsick2018}, XTE J1752--223 \citep{Garcia2018}, GX~339--4 \citep{Garcia2019}, MAXI~J1631--479 \citep{Xu2020}, XTE~J1908--094 \citep{Draghis2021}, MAXI~J1813--095 \citep{AJ2021a}. Most recently, \cite{Draghis2022} reported spins of ten new Galactic black hole candidate using the {\sc Relxill} model. 
Apart from these two, one can measure spin using timing properties where low-frequency Quasi Periodic Oscillations (QPOs) are assumed to originate from \textit{~Lense-Thirring} precession \citep{Ingram2009}.

\source~ or MAXI J1728--36 went into outburst on 17 January 2019 and continued until June 2019. The source was reported by Swift/BAT on 28 January 2019 \citep{Barthelmy2019a}. However, it was initially discovered by MAXI/GSC on 26 January 2019 \citep{Negoro19}. On 31 January 2019, MeerKAT detected the radio counterpart of the source with a magnitude of $11.2\pm0.6$ mJy at 1.28~GHz \citep{Bright2019}. Integral observed the source on 19 February 2019 and detected the source flux as $103\pm4$ mCrab in the $20-40$ keV range \citep{Ducci2019}. Following the MAXI and Swift detection, NICER has been monitoring the source since 29 January 2019 \citep{Enoto2019}, producing a total of 103 observations. Out of these, $\sim 70$ of them lie within the outbursting period. An optical counterpart with a magnitude of 16.7 was observed on 28 January 2019 by the 60~cm BOOTES-3/YA robotic telescope \citep{Hu2019}. Recently, \cite{Saha2022} analyzed NICER spectra of \source~ during the outburst and obtained a lower mass limit for the compact object of $4.6~M_{\odot}$, obtained assuming $a^* = 0$ from the inner radius of the disk and a distance of $\sim 10$ kpc. This estimate further strengthens \source's candidacy as a BHXRB. NuSTAR observed \source~ on 3 February 2019 but has thus far remained unexplored. This is the observation we utilized to estimate the spin and accretion properties.

In this paper, we concentrate on the MAXI/GSC light curve to explore the long time scale variability. We explored the first two observations of NICER to understand the beginning of the outburst. Later, we systematically examine the NuSTAR data to extract accretion disc properties and spin using numerous models such as \textsc{Diskbb}, \textsc{Cutoffpl}, \textsc{Nthcomp}, and \textsc{Relxill}. The paper is structured as follows. The data analysis process is briefly described in Section \ref{sec:obs}. Results obtained from MAXI/GSC light curve are presented in Section \ref{sec:delay}. Spectral analysis with NICER is presented in Section \ref{sec:NICER}. Spectral properties are analyzed and constrained in Section \ref{sec:analysis}, while {\sc Relxill} is employed in Section \ref{sec:Time_avg_rel}. Finally, we draw our conclusions and discuss our results in Section \ref{sec:conclusion}. All errors associated with model parameters are quoted at the 90\% confidence level unless otherwise stated.

\begin{table*}
\centering
\caption{Log of Observations of \source}
\begin{tabular}{lccccc}
\hline
Instrument & Date (UT) & Obs ID & Exposure (ks)  & Count s$^{-1}$\\
\hline \hline
NICER/XTI & 2019-01-29 & 1200550101 & 1.86 & $581.7 \pm 0.7$ \\ \hline
NICER/XTI & 2019-01-30 & 1200550102 & 13.7 & $992.0 \pm 0.4$ \\ \hline
NuSTAR & 2019-02-03 & 90501303002 & 55.9  & $144.7\pm0.02$\\ \hline

\end{tabular}
\label{tab:log}
\end{table*}

\section{Observations and Data Reduction}
\label{sec:obs}
NuSTAR observed \source~on 3 February 2019 for a total exposure of 55.9~ks (see Table~\ref{tab:log}). NuSTAR is a hard X-ray focusing telescope, consisting of two identical modules: FPMA and FPMB \citep{Harrison2013}. The raw data were reprocessed with the NuSTAR Data Analysis Software ({\tt NuSTARDAS}, version 1.4.1). Cleaned event files were generated and calibrated by using the standard filtering criteria in the {\tt nupipeline} task and the latest calibration data files available in the NuSTAR calibration database (CALDB)\footnote{\url{http://heasarc.gsfc.nasa.gov/FTP/caldb/data/nustar/fpm/}}. The source and background products were extracted by considering circular regions with radii 60 arcsec and 90 arcsec, at the source coordinates and away from the source, respectively. The spectra and light curves were extracted using the {\tt nuproduct} task. We re-binned the spectra with 25 counts per bin by using the {\tt grppha} task. Additionally, we divided the light-curves in two segments using {\tt xselect} and run {\tt nuproduct} using user defined {\tt gti} files. 

Following the discovery of \source, NICER began to monitor the source. We analysed the first two of the 70 NICER observations. NICER is a soft X-ray telescope whose primary X-ray Timing Instrument (XTI) consists of 56 identical FPMs (50 of which are functional) that record the energies of arriving photons, as well as their time of arrival to within 300 ns of UTC \citep{Gendreau2012}. Raw data were processed with the NICER Data Analysis Software ({\tt NICERDAS}, version 9)\footnote{\url{https://heasarc.gsfc.nasa.gov/docs/nicer/nicer_analysis.html}}. Cleaned event files were generated with standard filtering criteria in the {\tt nicerl2} task, while background files were generated with the {\tt nibackgen3C50} task. Response files were generated with the {\tt nicerarf} and {\tt nicerrmf} tasks. Spectra and light curves were extracted with {\tt xselect} and spectra were rebinned with 25 counts per bin using the {\tt grppha} task. We have not applied any systematic on NuSTAR or NICER data.

\section{Results and Discussion}
\subsection{Long Term Delay}
\label{sec:delay}

\begin{figure*}
\centering {
\includegraphics[width=1.0\textwidth]{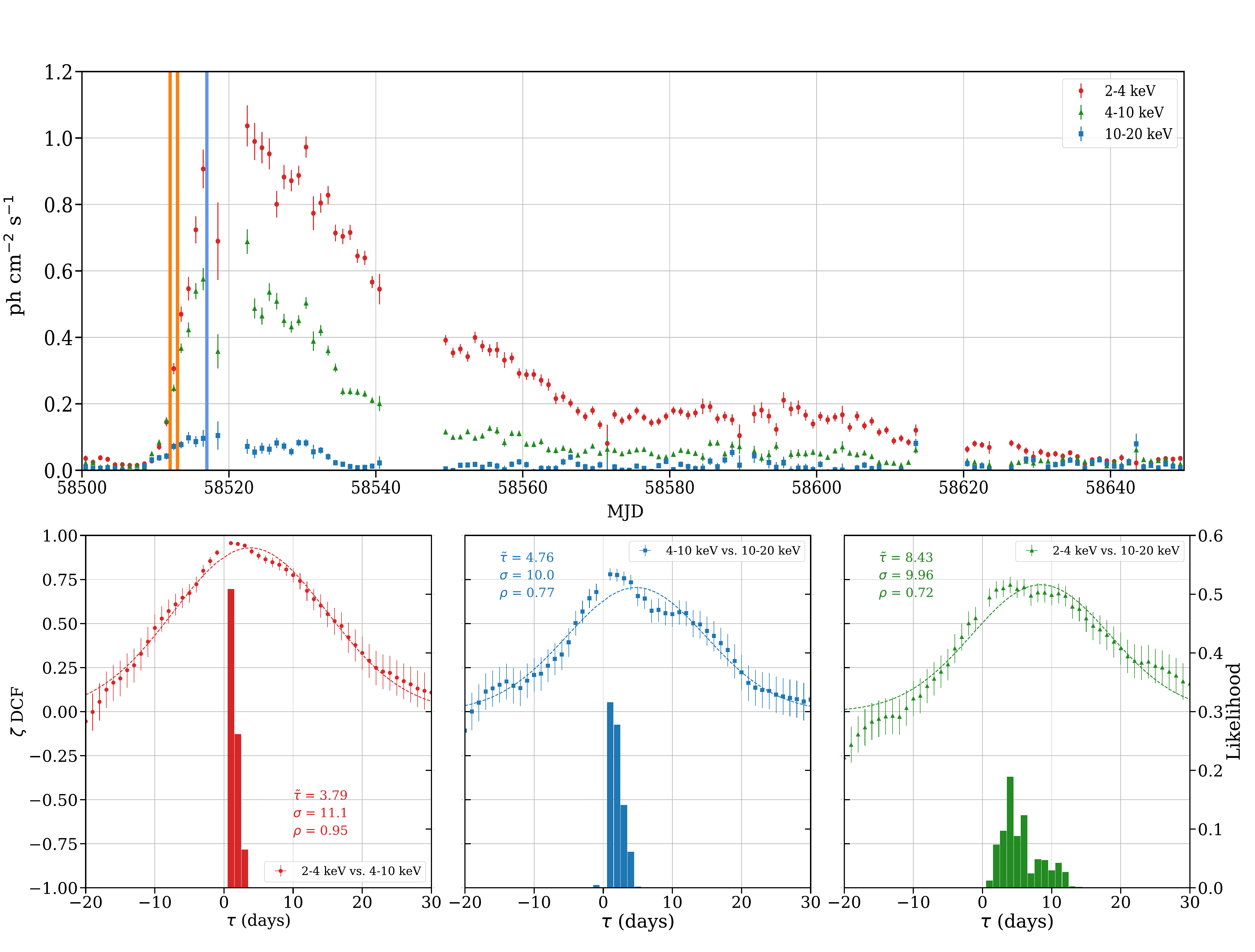}}
\caption{Top panel: MAXI/GSC light curves of \source~during the outburst in three different energy bands, binned at 1 day. The blue vertical line represents the date of the primary NuSTAR observation analyzed in Section \ref{sec:analysis}, while the orange lines show when the two NICER observations analyzed in Section \ref{sec:NICER} were taken. Bottom panels: Estimations of the cross--correlation functions between light curves of the indicated energy bands and likelihood distributions of the true time lags. Estimation was performed with the $\zeta$-transformed discrete correlation function ($\zeta$-DCF) and its associated likelihood (PLIKE) algorithm from \protect\cite{Tal2014}. $\protect\tilde{\tau}$ and $\sigma$ are the mean and standard deviation of the Gaussian fits, while $\rho$ is the maximum DCF value.}
\label{fig:lc}
\end{figure*}

MAXI/GSC \citep{Matsuoka2009} has monitored the source daily and provided one day binned light curves at various energy ranges. From these curves, as presented in the top panel of Fig. \ref{fig:lc}, we observed that the peak of the outburst occurred within 10-12 days of the beginning. Comparable to other canonical outbursts, rate variations were observed in various energy ranges. During the peak luminosity in the high/soft state, the hard 10-20 keV counts were substantially less than the softer 2-4 keV counts. In total, we found the outburst lasted for about 150 days starting from MJD 58500 to 58150.

The MAXI/GSC light curves can be used to obtain an estimate for the amount of energy released by the outburst in each energy band as
    $$E_i = 4\pi D^2 \tilde{E}_i \int_{{\rm MJD}_{\rm start}}^{{\rm MJD}_{\rm stop}} L_i(t) dt$$
where $D$ is the distance to the source, $\tilde{E}_i$ is the average photon energy in the $i^{\rm th}$ band, and $L_i$ is the corresponding MAXI/GSC light curve. Assuming a distance of 10 kpc and an average photon energy of 3.0 keV, 7.0 keV, and 15 keV in each energy band respectively, we calculate an energy output of $\sim 2.1 \times 10^{44}$ ergs in the $2-4$ keV band, $\sim 2.2 \times 10^{44}$ ergs in the $4-10$ keV band, and $\sim 9.0 \times 10^{43}$ ergs in the $10-20$ keV band, all over the course of the entire outburst. Considering all energy bands, the total energy released is $\sim 5.2 \times 10^{44}$ ergs, which translates to $5.8\times10^{26}$~g of mass assuming a mass-to-radiation conversion efficiency of 0.1 \citep{Reynolds2021}. The converted matter is equivalent to $2.9\times10^{-7} M_{\odot}$ or 91\% of the mass of Mars. However, the bolometric luminosity should be higher than what we have calculated from X-rays. Thus, it is expected that our estimation of the matter conversion is moderately reserved. 

\begin{figure*}
\centering{
\includegraphics[width=1.0\textwidth]{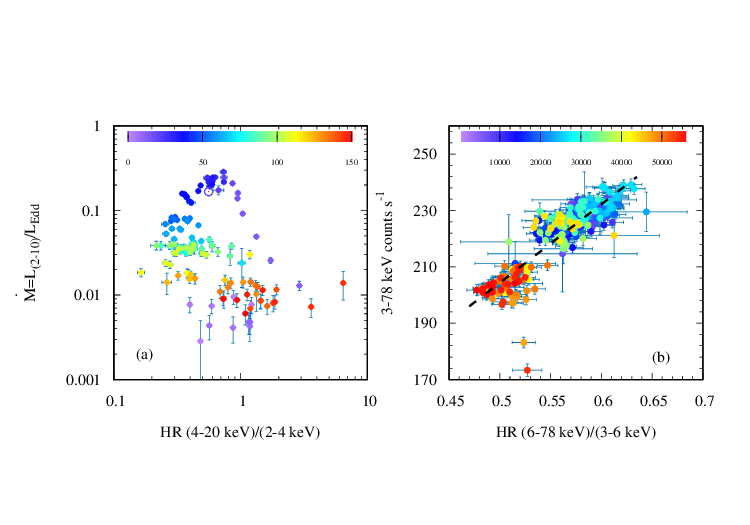}}
\caption{Left Panel (a): Hardness--accretion rate diagram of \source\ over the course of its outburst, produced with MAXI/GSC observations. The MAXI hardness ratio (HR) is defined as the ratio of count rates in the $4-20$ keV and $2-4$ keV bands. Colour represents progression of the outburst through time (MJD-58500 days) as indicated at the top. The maximum disc accretion rate was $\dot{M}_{\rm max}=0.28~L_{\rm Edd}$ during peak luminosity. The primary NuSTAR observation is marked with $\odot$. 
Right Panel (b): Hardness--intensity diagram (HID) for the first NuSTAR observation of the \source~outburst. Points are binned at 100s with the NuSTAR HR defined as the ratio of count rates in the $6-78$ keV and $3-6$ keV bands. Colour represents progression through time as indicated at the top, in seconds. The dashed line through the data is the best fit having a slope and intercept of $278.2\pm10.5$ and $65.2\pm6.0$ counts/s, respectively.}
\label{fig:HR1}
\end{figure*}

We inspected the time delays among the energy bands using the $\zeta$-DCF\footnote{\url{https://www.weizmann.ac.il/particle/tal/research-activities/software}} algorithm \citep{Tal2014}, presented in the bottom panels of Fig. \ref{fig:lc}. A positive delay refers to the softer component arriving later, while the reverse indicates delayed arrival of the harder component. The $2-4$ keV light curve correlates with both $4-10$ and $10-20$ keV light curves, having maximum correlation coefficients ($\rho$) of $0.95$ and $0.72$ respectively. The peak delays ($\tilde{\tau}$) between the corresponding light curves were found to be $(3.7 \pm 0.8)$ and $(8.4 \pm 1.9)$ days, with standard deviations $11.1$ and $9.9$ days respectively. A similar correlated pattern was also observed between $4-10$ and $10-20$ keV light curves with maximum coefficient, delay, and width $\rho=0.77$, $\tilde{\tau} = (4.7 \pm 1.1)$ days, $\sigma = 10.0$ days respectively. This delay is consistent with what we would expect examining the other two light curves, i.e. $8.4 \; {\rm days} - 3.7 \; {\rm days} = 4.7 \; {\rm days}$. We should note that $\sigma$ here refers to the estimated width of the cross-correlation function itself, whereas the errors associated with the true delays are estimated from the width of the corresponding likelihood distribution. These distributions were calculated with the PLIKE algorithm, also presented by \citet{Tal2014}. Our analysis showed that harder ($10-20$ keV) radiation as a whole arrived before its softer counterpart. The magnitude of the delay increased with increasing differences in X-ray energy, exhibiting a maximum delay between the $2-4$ and $10-20$ keV photons. An accreting BHXRB spectrum is typically approximated by the sum of a hard power-law and soft blackbody disk component, and from the light curve analysis we can infer that the harder spectral component dominated the rising phase before giving way to more prominent soft disk emission. This scenario is similar to what was observed earlier in the outburst profiles of GX 339--4, XTE J1650--500 \citep{Smith2002_accretion,Chatterjee2020}.  A possible reason for this delayed peak in soft emission could be the viscous delay with which {\bf the} standard disc spirals inward before heating enough to glow in the X-ray. Influenced by the viscous delay, the disc gradually modifies the spectral and temporal properties, triggering the outburst profile of the BHXRB \citep{Smith2002_accretion}. The left panel of Fig. \ref{fig:HR1} shows the variation in hardness ratio (HR) vs. intensity (in terms of the accretion rate $\dot M = L/L_{\rm Edd}$) over the full outburst as observed by MAXI/GSC, and illustrates a typical `q' diagram for an outbursting BHXRB.

\subsection{NICER}
\label{sec:NICER}
As mentioned previously, the X-ray spectrum of a BHXRB is typically approximated with the sum of a multi-colour disc blackbody (MCD) and power-law component. Reprocessed or reflected emission may also be observed, such as a Fe K$\alpha$ line at $\sim 6.4$~keV, a reflection hump at $\sim 15-40$~keV (see bottom panel of Fig. \ref{fig:nu-spec}), and in our case, an apparent Ni emission line at $\sim 8$~keV \citep{Corliss1981,Molendi2003,Medvedev2018}. Spectral analysis was carried out in {\tt HEASEARC}'s spectral analysis package {\tt XSPEC} version 12.12.1 \citep{Arnaud1996}. We used the {\sc Tbabs} model to account for interstellar absorption with the {\tt WILM} abundance \citep{Wilms2000} and the cross-section of \citet{Verner1996}. The MCD component was handled with the {\sc Diskbb} model, and is included in all spectral models presented in this paper.

The first two NICER observations of \source\ were taken on MJD 58512 and MJD 58513 respectively, which correspond to the beginning of the rising phase of the outburst -- or the low/hard state. We fit the data for the first two observations in the $0.3 - 10.0$ keV range with an absorbed MCD and power law model [{\sc Tbabs*(Diskbb + Powerlaw)}]. This returned an inner disc temperature of $T_{\rm in} = 1.05_{-0.02}^{+0.03}$ keV for the first observation and $T_{\rm in} = 1.229_{-0.006}^{+0.006}$ keV for the second. We find a reduction in photon index from $\Gamma = 2.02_{-0.07}^{+0.06}$ to $1.72_{-0.08}^{+0.08}$, with normalization varying from $\sim 1.7$ to $\sim 0.9~{\rm ph~keV^{-1}~cm^{-2}}$; hydrogen column density stayed approximately constant ($n_{\rm H} = 4.67_{-0.07}^{+0.07} \times 10^{-22}{\rm cm^{-2}}$ to $4.58_{-0.04}^{+0.04} \times 10^{-22} {\rm cm^{-2}}$). Table \ref{tab:nicer_fits} contains detailed results for each fit, along with their $\chi^2$ value per degree of freedom (dof) and model flux. Removing the MCD component from the model substantially worsens both fits, giving an F--statistic of $\sim 200$ and $\sim 3000$ for the first two observations respectively. This indicates the presence of a rapidly growing standard accretion disc at the beginning of the outburst.

\begin{table*}
    \centering
    \caption{NICER Spectral Analysis Results: {\sc Diskbb+Powerlaw}}
    \begin{tabular}{lccccccc}
    \hline
        Observation ID  & $n_{\rm H}$ ($10^{-22} {\rm cm^{-2}}$)  & $T_{\rm in}$ (keV)        & norm$_{\rm diskbb}$       & $\Gamma$                  & norm$_{\rm PL}$         & $\chi^2/$dof & $F_{2-10}$ ($10^{-9}$~ergs cm$^{-2}$) \\ \hline \hline
        1200550101      & $4.67_{-0.07}^{+0.07}$            & $1.05_{-0.02}^{+0.03}$    & $109_{-20}^{+20}$ & $2.02_{-0.07}^{+0.06}$    & $1.7_{-0.3}^{+0.3}$   & 891/902 & $4.498_{-0.010}^{+0.010}$ \\ \hline
        1200550102      & $4.58_{-0.04}^{+0.04}$            & $1.229_{-0.006}^{+0.006}$ & $215_{-8}^{+8}$   & $1.72_{-0.08}^{+0.08}$    & $0.9_{-0.2}^{+0.2}$   & 1008/944 & $6.177_{-0.014}^{+0.014}$ \\ \hline
    \end{tabular}
    \leftline{$F_{2-10}$ is the total model flux in the $2-10$ keV range.}
    \label{tab:nicer_fits}
\end{table*}

\subsection{NuSTAR}
\label{sec:analysis}

\begin{figure*}
\centering {
\includegraphics[width=\textwidth]{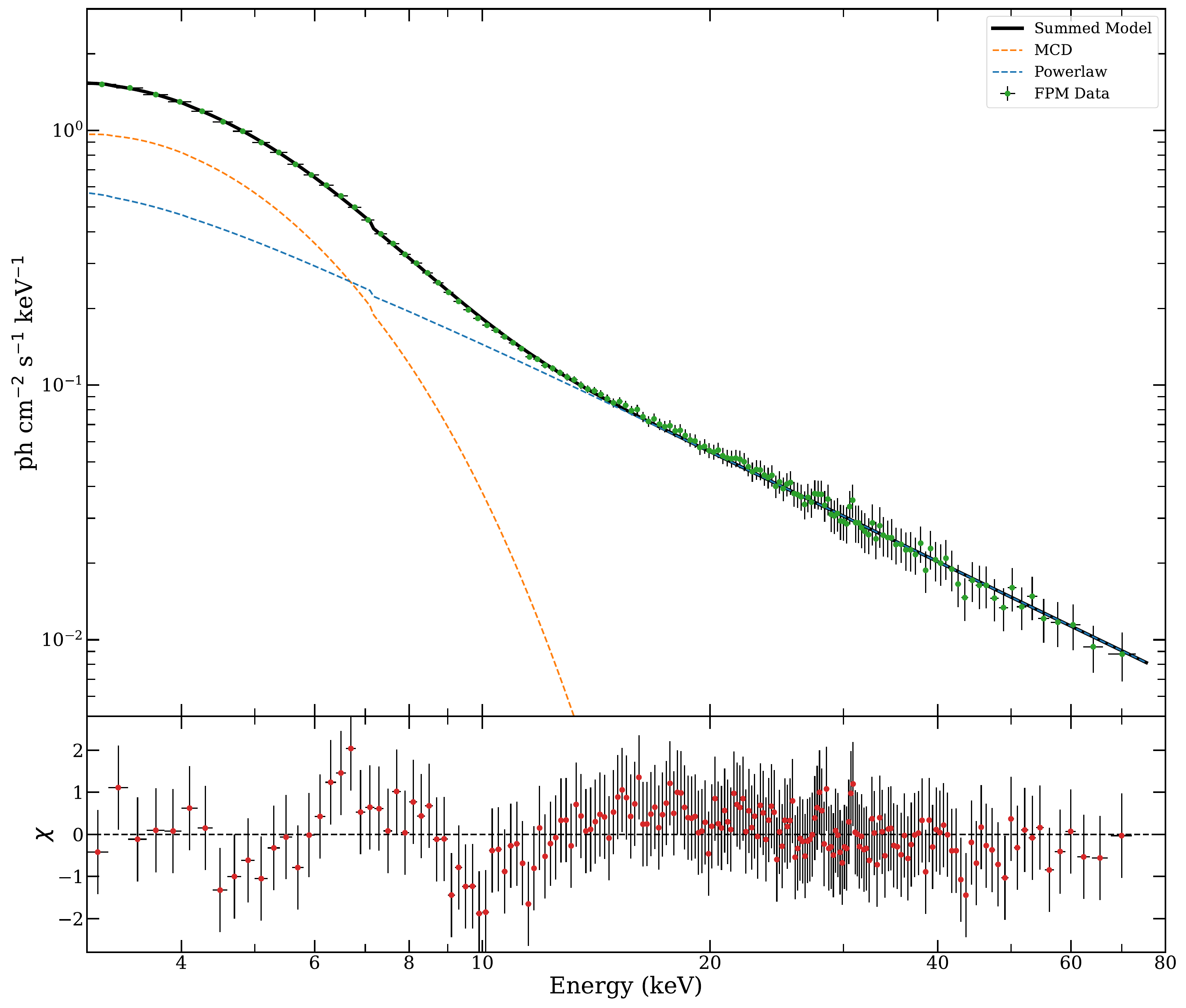}}
\caption{Top panel: Time-averaged unfolded spectrum from FPMB (green) fit with an absorbed MCD and power law model, with no Gaussian components. Both the individual model components as well as their sum are presented. Bottom Panel: Residuals in terms of (data$-$model)/error obtained from spectral analysis for the same model.}
\label{fig:nu-spec}
\end{figure*}

We performed the spectral analysis of the NuSTAR data in the $3-78$~keV energy range. The right panel of Fig.~\ref{fig:HR1} shows the variation in HR vs. count rate throughout the primary NuSTAR observation, with points binned at 100s. We observe that during the observation HR and intensity were positively correlated, lying along the line with slope, intercept, correlation coefficient, and null hypothesis probability $(278.2\pm10.5)$, $(65.2\pm6.0)$ counts/s, $r =0.877$, and $p<10^{-5}$ respectively. Correlation between these two parameters is typical of BHXRBs during intermediate spectral states where disc-corona coupling becomes
stronger \citep{Churazov2001,Taylor2003}. In Section \ref{sec:HR_spec} we resolve the spectrum between the two regions in the HR diagram with unfolded spectra from both regions presented in Fig. \ref{fig:hr-spec}. Section \ref{sec:Time_avg_ph} is dedicated to examining the full time-averaged spectrum of the observation using several non-relativistic models, and in Section \ref{sec:Time_avg_rel} we explore the same spectrum with several flavours of the {\sc Relxill} model family. Gaussian components are included to model the Fe and Ni emission lines where necessary in all non-relativistic models. Fig. \ref{fig:nu-spec} shows the full spectrum of the NuSTAR observation fit with a simple MCD and power-law model, which disregards all reflected emission. The bottom panel of Fig. \ref{fig:nu-spec} displays the residuals of that fit and showcases the reflected emission, i.e. the 6.4 keV Fe K$\alpha$ line and reflection hump above $\sim 15$ keV.  

 \subsubsection{Hardness-Resolved Spectra}
 \label{sec:HR_spec}
\begin{figure*}
\centering {
\includegraphics[width=\textwidth]{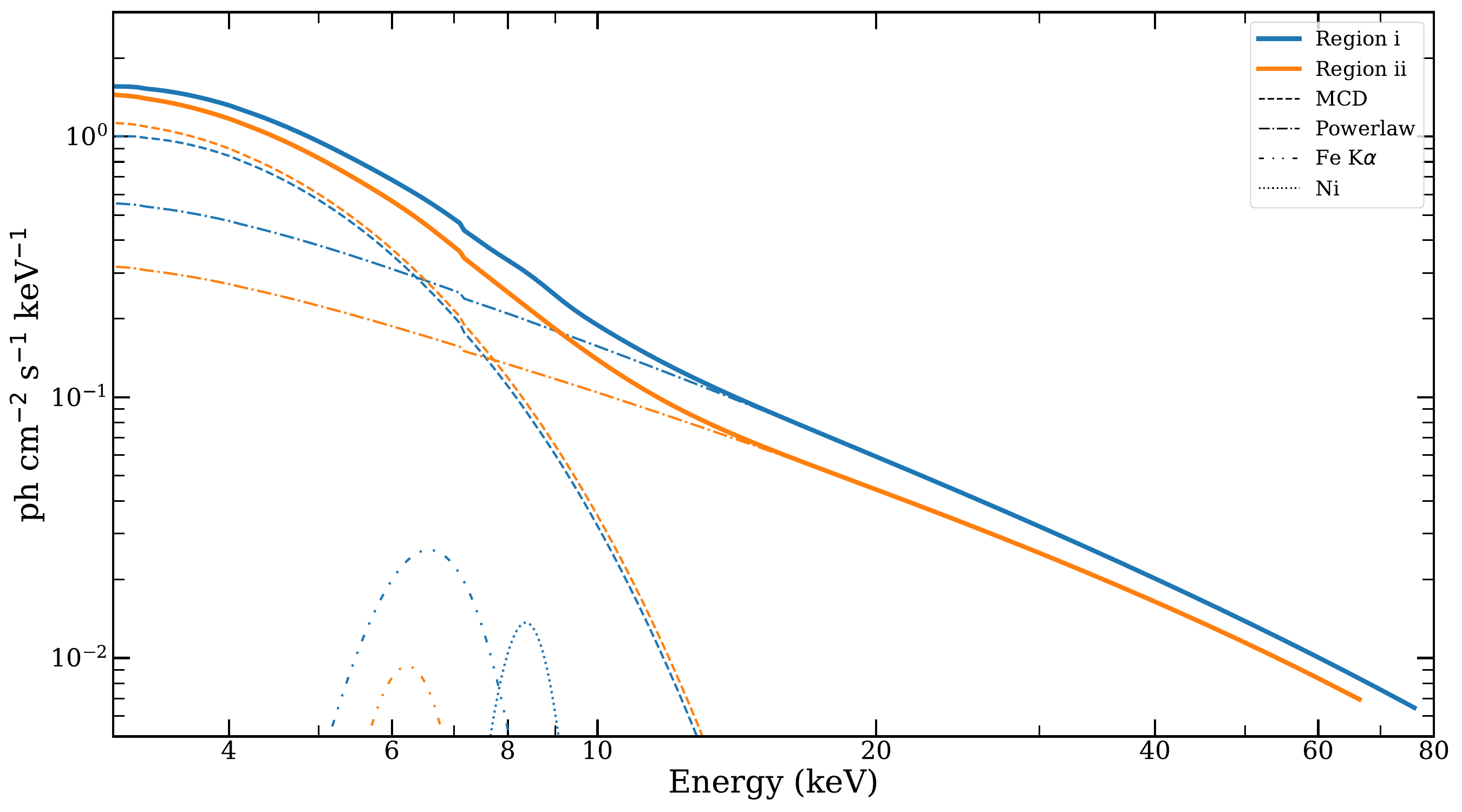}}
\caption{Models (both total and their individual components) for both spectra extracted from region i and region ii of the HID presented in the right panel of Fig. \ref{fig:HR1}.}
\label{fig:hr-spec}
\end{figure*}

The `gap' between data points that occurs at $\sim 45,000$s shows a sudden softening of \source's spectrum, alongside a reduction in intensity during the observation. We extracted spectra for both regions of the HID (denoting the earlier region i and the later region ii) and fit them with an absorbed MCD and cutoff power law model {\sc Cutoffpl}, which very roughly approximates the continuum emission of a thermally Comptonized medium. Fig. \ref{fig:hr-spec} shows the resulting spectral fits for both regions.

Between regions i and ii the inner disc temperature, photon index, and cutoff energy did not appreciably vary, returning values of $T_{\rm in} = 1.195_{-0.011}^{+0.010}$ keV, $\Gamma = 2.27_{-0.07}^{+0.07}$, and $E_{\rm cut} = 94_{-34}^{+78}$ keV in region i; and $T_{\rm in} = 1.216_{-0.005}^{+0.005}$ keV, $\Gamma = 2.13_{-0.03}^{+0.03}$, and $E_{\rm cut} = 79_{-18}^{+28}$ keV in region ii. In region i we also detected signatures of both the Fe K$\alpha$ line at $E_{\rm Fe} = 6.34_{-0.14}^{+0.11}$ keV and the Ni {\sc XXVIII} line at $E_{\rm Ni} = 8.06_{-0.28}^{+0.10}$ keV. 
In region ii we only detected the Fe K$\alpha$ line (with greater uncertainty) at $E_{\rm Fe} = 6.2_{-0.3}^{+0.4}$ keV, and the apparent hydrogen column density changed from $n_{\rm H} = 4.19_{-0.12}^{+0.25} \times 10^{-22}{\rm cm^{-2}}$ to a more uncertain $n_{\rm H} = 3.14_{-0.7}^{+0.7} \times 10^{-22} {\rm cm^{-2}}$. Between regions we found a reduction in the harder power law component, with normalization varying from $\sim 3.3$ to $\sim 1.5~{\rm ph~keV^{-1}~cm^{-2}}$ (illustrated in Fig. \ref{fig:hr-spec}). This explains the softening of the spectrum as well as the disappearance of the reprocessed Ni emission line. Variations on this timescale ($\sim 100$s) being primarily influenced by the power-law component are consistent with the findings of \citet{Churazov2001} in the soft state of Cyg X-1. Detailed results of both fits are presented in Table \ref{tab:split_spectra}.

{\renewcommand{\arraystretch}{1.2}
\begin{table}
     \centering
     \caption{NuSTAR Hardness Resolved Spectral Results}
     \begin{tabular}{lcc}
    \hline
        & i & ii \\ \hline \hline
        $n_{\rm H}$ ($10^{22} {\rm cm^{-2}}$) & $4.19_{-0.12}^{+0.25}$ & $3.8_{-0.7}^{+0.7}$ \\
        $T_{\rm in}$ (keV) & $1.195_{-0.011}^{+0.010}$ & $1.216_{-0.005}^{+0.005}$ \\
        norm$_{\rm diskbb}$ & $313_{-12}^{+13}$ & $303_{-6}^{+7}$ \\
        $\Gamma$ & $2.27_{-0.07}^{+0.07}$ & $2.13_{-0.03}^{+0.03}$ \\
        $kT_e$ (keV) & $94_{-34}^{+78}$ & $79_{-18}^{+28}$ \\
        norm$_{\rm PL}$ (${\rm ph~keV^{-1}cm^{-2}}$) & $3.3_{-0.5}^{+0.6}$ & $1.54_{-0.07}^{+0.07}$ \\
        $E_{\rm Fe}$ (keV) & $6.34_{-0.14}^{+0.11}$ & $6.2_{-0.3}^{+0.4}$ \\
        $\sigma_{\rm Fe}$ (keV) & $0.80_{-0.07}^{+0.08}$ & $0.3_{-0.3}^{+0.5}$ \\
        norm$_{\rm Fe}$ ($10^{-3}$) & $6.5_{-1.4}^{+0.5}$ & $1.1_{-0.9}^{+1.0}$ \\
        $E_{\rm Ni}$ (keV) & $8.06_{-0.28}^{+0.10}$ & - \\
        $\sigma_{\rm Ni}$ (keV) & $0.37_{-0.13}^{+0.19}$ & - \\
        norm$_{\rm Ni}$ ($10^{-3}$) & $1.7_{-0.3}^{+1.2}$ & - \\ \hline
        $\chi^2$/dof & 869/887 & 604/599 \\
        $F_{2-10}$ ($10^{-9}$~ergs cm$^{-2}$) & $9.952_{-0.012}^{+0.012}$ & $8.862_{-0.022}^{+0.022}$ \\\hline \hline
     \end{tabular}
     \label{tab:split_spectra}
\end{table}
}

\begin{figure*}
\centering {
\includegraphics[width=\textwidth]{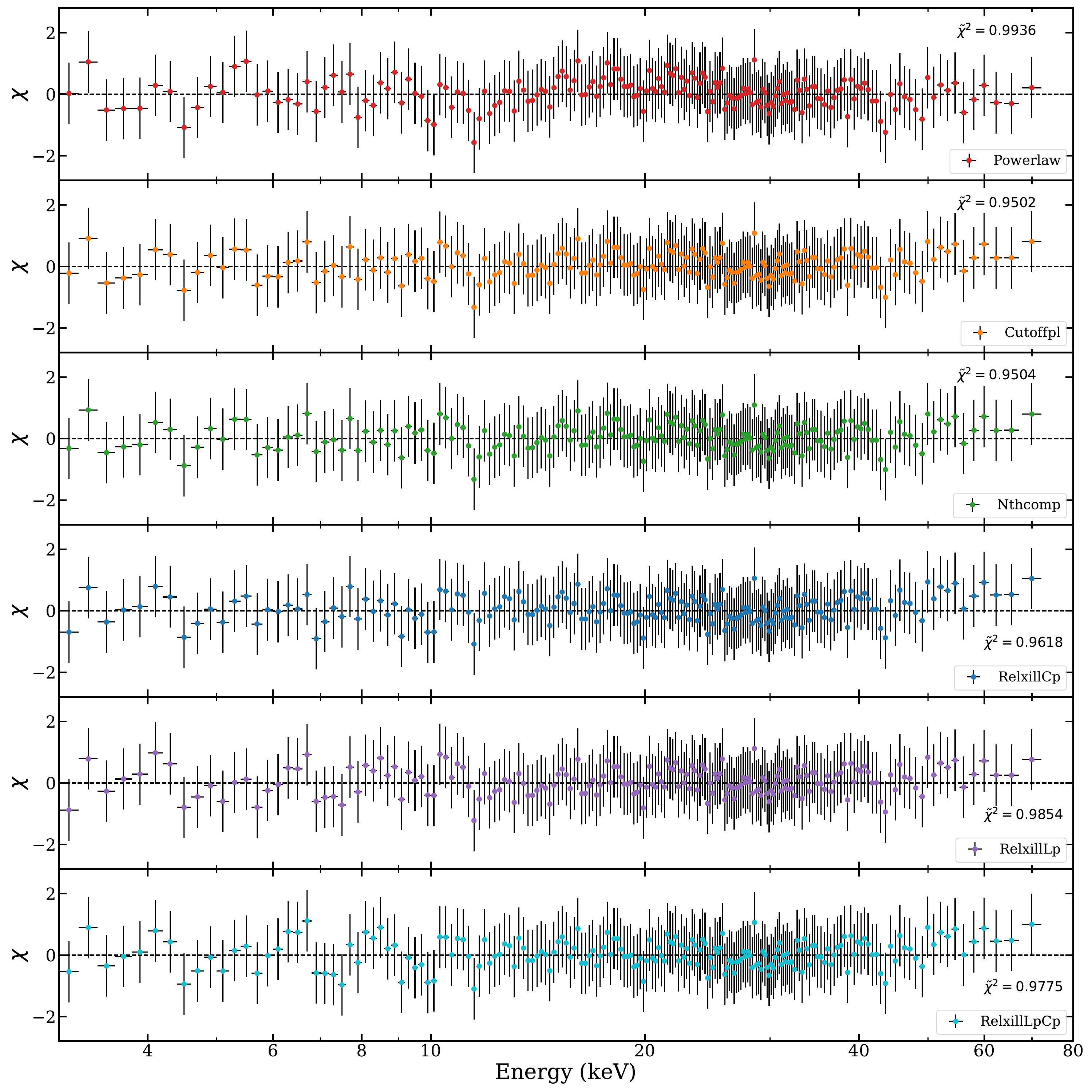}}
\caption{Residuals in terms of (data-model)/error obtained from spectral analysis for the {\sc Powerlaw}, {\sc Cutoffpl}, {\sc Nthcomp}, {\sc RelxillCp}, {\sc RelxillLp}, and {\sc RelxillLpCp} models. Reduced $\chi^2$ values ($\tilde{\chi}^2$) for each model are quoted in each panel. The first three models (i.e., the non-relativistic ones) include two {\sc Gaussian} components for the Fe K$\alpha$ and Ni emission lines.}
\label{fig:del}
\end{figure*}

\subsubsection{Time averaged spectrum: Phenomenological Models}
 \label{sec:Time_avg_ph}
 
Examining the entire spectrum of the observation, we first fit the data with an absorbed power--law and MCD model. This model returned a reasonable fit with $\chi^2 = 950$ for 956 degrees of freedom (dof) with estimates of the inner disc temperature $T_{\rm in} = 1.185_{-0.010}^{+0.015}$ keV, a photon index of $\Gamma = 2.46_{-0.02}^{+0.02}$, and a column density of $n_{\rm H} = 4.63_{-0.28}^{+0.28} \times 10^{-22}\mathrm{cm^{-2}}$. We find signatures of the Fe and Ni emission lines at $E_{\rm Fe} = 6.48_{-0.02}^{+0.02}$ keV and $E_{\rm Ni} = 7.99_{-0.03}^{+0.04}$ keV. The reflection hump was clearly visible in the residuals at energies above $\sim 15$~keV (see Fig. \ref{fig:del}). This was repeated with an exponential cutoff power-law component {\sc Cutoffpl}. This provided no significant improvement to the fit, returning an inner disc temperature of $T_{\rm in} = 1.16_{-0.03}^{+0.03}$ keV, a slightly softer photon index of $\Gamma = 2.24_{-0.08}^{+0.08}$, cutoff energy $kT_e = 94_{-39}^{+93}$ keV, and column density $n_{\rm H} = 4.32_{-0.23}^{+0.27} \times 10^{-22}\mathrm{cm^{-2}}$; we again find signatures of the Fe and Ni emission lines at $E_{\rm Fe} = 6.27_{-0.13}^{+0.16}$ keV and $E_{\rm Ni} = 8.0_{-0.2}^{+0.3}$ keV. The reflection hump was still clearly visible in the residuals. 

A much better description of continuum emission due to thermal Comptonization is given by the model \textsc{Nthcomp} \citep{Z96}, which attempts to simulate the upscattering of photons through the corona from a seed spectrum parameterized by the inner temperature of the accretion disc. The high energy cutoff is parameterized by the electron temperature of the medium. {\sc Nthcomp} is not a power law, but can still be parameterized with an asymptotic photon index \citep{Zycki1999}. Using the inner disc temperature returned by {\sc Diskbb}, $T_{\rm in} = 1.149_{-0.009}^{+0.006}$ keV, as the low energy rollover in {\sc Nthcomp}, we estimate the hot electron temperature to be $kT_e = 132_{-123}^{+476}$ keV with photon index $\Gamma = 2.36_{-0.01}^{+0.02}$. This model also returned the lowest hydrogen column density of all of the presented models with $n_{\rm H} = 3.27_{-0.35}^{+0.91} \times 10^{-22}\mathrm{cm^{-2}}$. For completeness, we may also calculate the optical depth ($\tau$) of the corona as given by \cite{Z96},
    $$\tau = \sqrt{\frac{9}{4} + \frac{m_e c^2}{kT_e}\frac{3}{(\Gamma - 1)(\Gamma +2)}} - \frac{3}{2},$$
returning a value of $\tau = 0.55_{-0.42}^{+3.54}$ for the {\sc Nthcomp} model. The optically thin corona was also proposed for Cyg X-1 \citep{Churazov2001} during its high/soft state.

We also attempted to fit the spectrum with the non-relativistic reflected power-law model {\sc Pexrav} from \cite{MZ1995}, alongside a regular power-law component to account for direct coronal emission. Since this model accounts for the reflection and reprocessing of X-rays from the corona by the accretion disc, it can be used to estimate the reflection fraction ($R_{\rm refl}$), defined as the fraction of X-rays reprocessed by the disc vs. those received via direct emission. It also estimates abundances of elements heavier than He, including iron ($A_\mathrm{Fe}$), as well as the inclination angle $i$. This model returned a close fit with the reflection hump reduced in the residuals and a $\chi^2 = 916$ for 949 dof. However, most parameters were unable to be constrained. Hence, we opted for relativistic model to estimate those parameters. One of the few parameters that remained well-behaved was the reflection fraction, $R_{\rm refl} = 0.48_{-0.06}^{+0.06}$, and we should expect {\sc Relxill} to estimate reflection fractions of the same order. 

Finally, we should note explicitly that in all the aforementioned models, two {\sc Gaussian} model components were used to model to the Fe K$\alpha$ and Ni emission lines. We observed two {\bf more} emission lines at 27.9~keV and 30.8~keV having widths of 0.07~keV and 0.06~keV respectively; the normalization of both lines was around $10^{-4}$~ph cm$^{-2}$ s$^{-1}$. The origin of these emission lines could be attributed to the radioactive isotope $^{241}$Am, as reported by \cite{Garcia2018,Connors2022}. Similar lines were also observed in AstroSat/LAXPC spectra, which are an instrumental feature \citep{Sreehari2019}. To fit those lines, we added two additional {\sc Gaussian} components to each model and froze the line energy, width, and normalization as obtained from the {\sc Powerlaw} model. Detailed results of all spectral fits for this section are presented in the first three columns of Table \ref{tab:spec_results}, and the residuals for each model are presented in Fig. \ref{fig:del}.

\subsubsection{Time averaged spectrum: Relxill}
\label{sec:Time_avg_rel}
We used different flavors of the relativistic reflection model \textsc{Relxill} \citep{Garcia2013,Garcia2014,Dauser2014,Dauser2016} to probe the reprocessed emission more accurately. In this model, the reflection fraction ($R_{\rm refl}$) is given by the ratio between the Comptonized emission to the disc and to the infinity. A broken power-law emission profile is assumed with $E(r) \sim R^{-q_{\rm in}}$ for $r > R_{\rm br}$ and $E(r) \sim R^{-q_{\rm out}}$ for $r < R_{\rm br}$, where E(r), $q_{\rm in}$, $q_{\rm out}$ and $R_{\rm br}$ are the emissivity, inner emissivity index, outer emissivity index, and break radius respectively. The other free parameters in this model are the ionization parameter ($\xi$), iron abundance ($A_{\rm Fe}$) and inclination angle ($i$).

We started our analysis with the \textsc{RelxillCp} model, with the final model reading in \texttt{XSPEC} as \textsc{TBabs*(diksbb+RelxillCp)}. The primary emission for this model is given by the Comptonized model \textsc{Nthcomp} \citep{Z96,Zycki1999}. To verify that {\sc Relxill} does indeed present an advantage over our non-relativistic models, we compared {\sc Nthcomp} with no additional {\sc Gaussian} components to model the Fe/Ni emission lines to the {\sc RelxillCp} model through an F-test, returning an F-statistic of $\sim 15$ (probability $3.1 \times 10^{-25}$). This confirms a statistical need to include modeling of reflected emission in the NuSTAR spectrum. \textsc{RelxillCp} directly estimates the coronal properties in terms of the $\Gamma$ and $kT_{\rm e}$. During fitting, we linked the seed photon temperature ($T_{\rm S}$) with the inner disc temperature ($T_{\rm in}$). We fixed the outer radius of the disc at $R_{\rm out}=1000$ $r_g$. 

Analysis with \textsc{RelxillCp} returned a good fit with $\chi^2=916$ for 952 dof. We obtained an inner temperature of $T_{\rm in} = 1.2358_{-0.0007}^{+0.0007}$ keV, photon index $\Gamma = 2.381_{-0.003}^{+0.003}$,  iron abundance $A_{\rm Fe} = 0.47_{-0.06}^{+0.06}~A_{\odot}$, and ionization parameter $\log(\xi) = 3.39_{-0.04}^{+0.04}$. The ISCO was extended up to $R_{\rm in} = 2.2_{-0.6}^{+0.4}$~$R_{\mathrm{ISCO}}$, with the break radius extending out to $R_{\rm br} = 14.3_{-0.6}^{+0.6}$ $r_g$. The inner emissivity index appears steep with $q_1=8.6_{-0.4}^{+0.4}$, and we find a much flatter outer emissivity index of $q_2=0.35_{-0.11}^{+0.10}$. We estimate a reflection fraction of $R_{\rm refl} = 0.67_{-0.06}^{+0.06}$ and a BH spin parameter of $a^* = 0.65_{-0.06}^{+0.04}$, with inclination angle $i = 69.8_{-0.5}^{+0.5}$ degrees.

While the {\sc RelxillCp} model does not assume any particular coronal geometry, the {\sc RelxillLp} and {\sc RelxillLpCp} flavors of the \textsc{Relxill} family assume a lamp-post geometry where the corona is assumed to be a compact source located above the BH \citep{Garcia2010,Dauser2016}. The incident primary emission is given by either \textsc{Cutoffpl} (\textsc{RelxillLp}) or \textsc{Nthcomp} (\textsc{RelxillLpCp}). The height of corona ($h$) is an input parameter in this model.

Analysis with the {\sc RelxillLp} and {\sc RelxillLpCp} models both returned good fits with $\chi^2/{\rm dof} = 940/954$ and $\chi^2 / {\rm dof} = 933/954$ respectively. Both returned similar inner temperatures of $T_{\rm in} = 1.220_{-0.004}^{+0.004}$ keV and $T_{\rm in} = 1.211_{-0.005}^{+0.005}$ keV respectively. {\sc RelxillLp} returned a photon index of $\Gamma = 2.421_{-0.003}^{+0.003}$ while {\sc RelxillLpCp} returned a slightly harder photon index of $\Gamma = 2.244_{-0.002}^{+0.002}$. {\sc RelxillLp} also returned a reflection fraction of $R_{\rm refl} = 0.76_{-0.04}^{+0.04}$, almost two times more than the $R_{\rm refl} = 0.470_{-0.014}^{+0.014}$ given by {\sc RelxillLpCp}. Both models returned similar values of the coronal height, with $h = 3.59_{-0.11}^{+0.15}~r_g$ for {\sc RelxillLp} and $h = 5.484_{-0.009}^{+0.009}~r_g$ for {\rm RelxillLpCp}. Otherwise, {\sc RelxillLp} returned an iron abundance $A_{\rm Fe} = 1.50_{-0.12}^{+0.12}~A_{\odot}$, ionization parameter $\log(\xi) = 4.48_{-0.05}^{+0.05}$, $R_{\rm in} = 2.5_{-0.5}^{+0.5}~R_{\mathrm{ISCO}}$, inclination angle $i = 58_{-2}^{+2}$ degrees, and spin parameter $a^* = 0.70_{-0.31}^{+0.15}$. {\rm RelxillLpCp} returned an iron abundance $A_{\rm Fe} = 0.44_{-0.04}^{+0.04}~A_{\odot}$, ionization parameter $\log(\xi) = 3.01_{-0.02}^{+0.02}$,  $R_{\rm in} = 1.7_{-0.5}^{+1.4}~R_{\mathrm{ISCO}}$, inclination angle $i = 37_{-3}^{+3}~^{\circ}$, and spin parameter $a^* = 0.63_{-0.07}^{+0.07}$. Detailed results of all spectral fits for this section are presented in the last three columns of Table \ref{tab:spec_results}.

\subsection{Error Estimation}
\begin{figure*}
\centering
\includegraphics[height=0.9\textwidth,width=\textwidth]{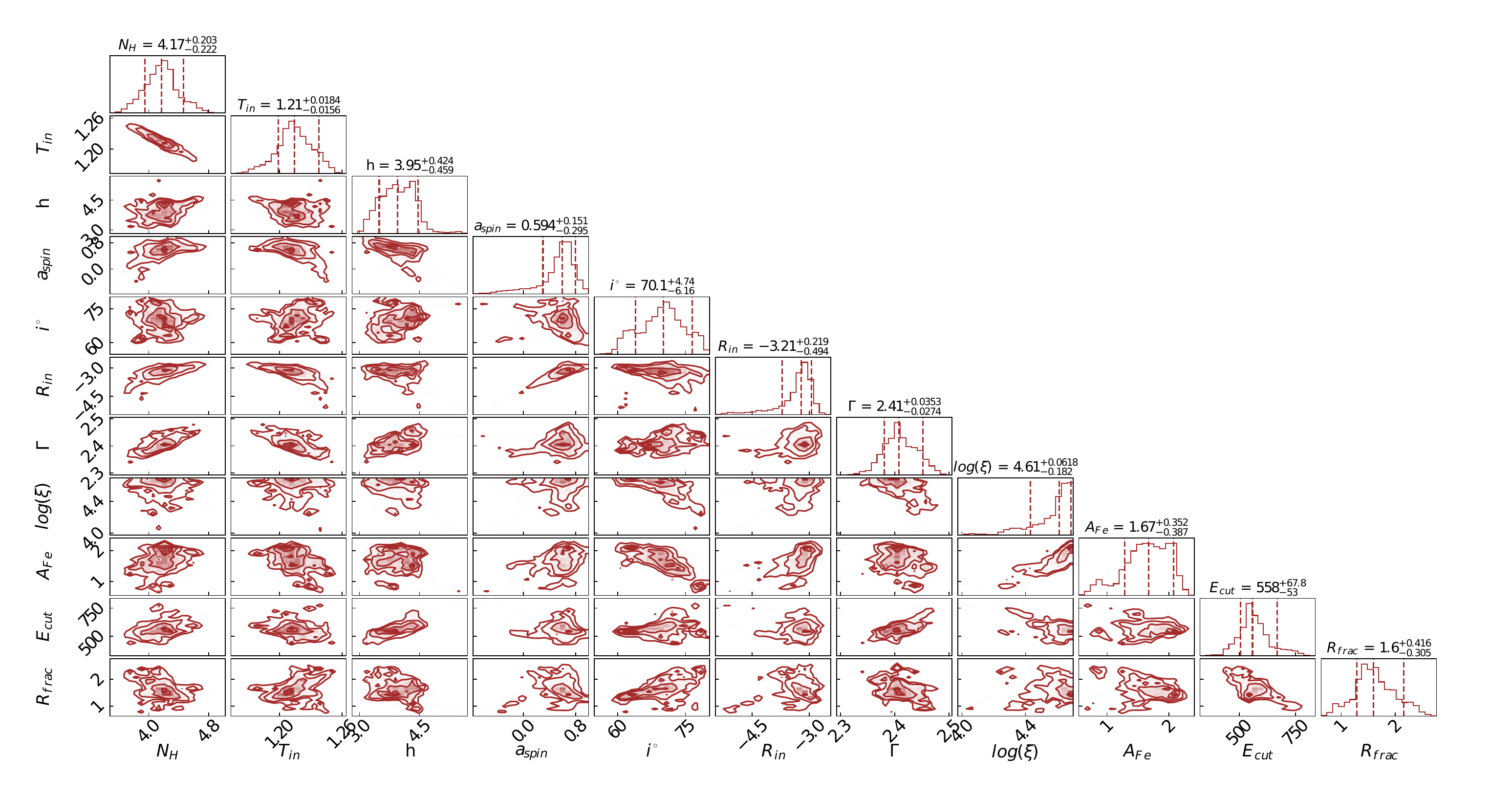}
\caption{Posterior distribution of the spectral parameters obtained from the MCMC analysis with the \textsc{RelxillLp} model. Plotting was performed using \texttt{corner} \citep{corner}. Central dashed lines correspond to the peak values whereas $1\sigma$ confidence levels are represented by dashed lines on either side.}
\label{fig:mcmc}
\end{figure*}

While fitting the NuSTAR spectrum, we found that some of the parameters were degenerate. To remove this degeneracy and find the global minimum, we employed Markov Chain Monte Carlo (MCMC) in {\tt XSPEC}\footnote{\url{https://heasarc.gsfc.nasa.gov/xanadu/xspec/manual/node43.html}}, which constrains the uncertainty range of the parameters. We used the \textsc{RelxillLp} model to estimate the errors using MCMC, as the model provides a physical picture of the corona. As a caveat, it should be noted that the coronal geometry influences the emissivity profile of the reflection spectrum. Thus, the parameters could modify if coronal geometry varies. We used 1,000,000 steps with 8 walkers using the Goodman-Weare algorithm. We chose to discard the first 10,000 steps, i.e. the `transient' or `burn-in' phase. The posterior distribution of the \textsc{RelxillLp} model fitted parameters are plotted in Figure~\ref{fig:mcmc}, with errors quoted at the 1 $\sigma$.

{\renewcommand{\arraystretch}{1.2}
\begin{table*}
\centering
\caption{NuSTAR Spectral Analysis Results}
\label{tab:spec_results}
\begin{tabular}{lccccccc}
\hline
                                                 & Powerlaw                  & Cutoffpl               & Nthcomp                   & RelxillCp              & RelxillLp                 & RelxillLpCp       \\ \hline \hline
$n_\mathrm{H}$ ($10^{-22}\mathrm{cm^{-2}}$)     & $4.63_{-0.28}^{+0.28}$    & $4.32_{-0.23}^{+0.27}$ & $3.27_{-0.35}^{+0.91}$    & $4.05_{-0.03}^{+0.03}$  & $4.12_{-0.09}^{+0.09}$    & $4.32_{-0.11}^{+0.11}$   \\
$T_\mathrm{in}$ (keV)                           & $1.185_{-0.010}^{+0.015}$ & $1.16_{-0.03}^{+0.03}$ & $1.149_{-0.009}^{+0.006}$ & $1.2358_{-0.0007}^{+0.0007}$   & $1.220_{-0.004}^{+0.004}$ & $1.211_{-0.005}^{+0.005}$   \\
$\Gamma$                                        & $2.46_{-0.02}^{+0.02}$    & $2.24_{-0.08}^{+0.08}$ & $2.36_{-0.01}^{+0.02}$    & $2.381_{-0.003}^{+0.003}$   & $2.421_{-0.004}^{+0.004}$    & $2.244_{-0.002}^{+0.002}$   \\
$E_\mathrm{Fe}$ (keV)                           & $6.48_{-0.02}^{+0.02}$    & $6.27_{-0.13}^{+0.16}$ & $6.40_{-0.02}^{+0.12}$    & -                 & -                         & -                 \\
$\sigma_\mathrm{Fe}$ (keV)                      & $0.50_{-0.05}^{+0.05}$    & $0.6_{-0.2}^{+0.3}$    & $0.62_{-0.07}^{+0.10}$    & -                 & -                         & -                 \\
norm$_\mathrm{Fe}$ ($10^{-3}$)                  & $3.5_{-0.8}^{+0.4}$       & $4.6_{-0.2}^{+5.3}$    & $6.0_{-0.8}^{+1.2}$       & -                 & -                         & -                 \\
$E_\mathrm{Ni}$ (keV)                           & $7.99_{-0.03}^{+0.04}$    & $8.0_{-0.2}^{+0.3}$    & $8.1_{-0.1}^{+0.2}$       & -                 & -                         & -                 \\
$\sigma_\mathrm{Ni}$ (keV)                      & $0.56_{-0.06}^{+0.07}$    & $0.5_{-0.2}^{+0.3}$    & $0.48_{-0.18}^{+0.12}$    & -                 & -                         & -                 \\
norm$_\mathrm{Ni}$ ($10^{-3}$)                  & $1.7_{-0.6}^{+0.4}$       & $1.3_{-1.3}^{+0.9}$    & $1.7_{-1.1}^{+1.0}$       & -                 & -                         & -                 \\
$kT_e$ (keV)                                    & -                         & $94_{-39}^{+93}$       & $132_{-123}^{+476}$       & $400^f$      & $400^f$       & $400^f$   \\
$R_\mathrm{refl}$                               & -                         & -                      & -                         & $0.67_{-0.06}^{+0.06}$  & $0.76_{-0.04}^{+0.04}$    & $0.470_{-0.014}^{+0.014}$  \\
$A_\mathrm{Fe}$ ($A_{\sun}$)                    & -                         & -                      & -                         & $0.47_{-0.06}^{+0.06}$  & $1.50_{-0.12}^{+0.12}$    & $0.44_{-0.04}^{+0.04}$  \\
$a^*$                                           & -                         & -                      & -                         & $0.65_{-0.06}^{+0.04}$  & $0.70_{-0.31}^{+0.15}$    & $0.63_{-0.07}^{+0.07}$  \\
$i$ ($^\circ$)                                  & -                         & -                      & -                         & $69.8_{-0.5}^{+0.5}$   & $58_{-2}^{+2}$            & $37_{-3}^{+3}$   \\
$R_\mathrm{in}$ ($R_{\mathrm{ISCO}}$)                         & -                         & -                      & -                         & $2.2_{-0.6}^{+0.4}$   & $2.5_{-0.5}^{+0.5}$       & $1.7_{-0.5}^{+1.4}$   \\
$R_\mathrm{br}/h$ ($r_g$)                       & -                         & -                      & -                         & $14.3_{-0.6}^{+0.6}$   & $3.59_{-0.11}^{+0.15}$    & $5.484_{-0.009}^{+0.009}$  \\
$\log\xi$                                       & -                         & -                      & -                         & $3.39_{-0.04}^{+0.04}$   & $4.48_{-0.05}^{+0.05}$          & $3.01_{-0.02}^{+0.02}$   \\
$q_1$                                           & -                         & -                      & -                         & $8.61_{-0.41}^{+0.37}$   & -                         & -                 \\
$q_2$                                           & -                         & -                      & -                         & $0.35_{-0.11}^{+0.10}$  & -                         & -                 \\
norm$_{\rm diskbb}$                             & $305_{-21}^{+14}$         & $379_{-31}^{+33}$      & $464_{-16}^{+21}$         & $276_{-6}^{+6}$     & $265_{-6}^{+6}$         & $299_{-8}^{+8}$     \\
norm                                            & $4.5_{-0.2}^{+0.2}$       & $2.9_{-0.5}^{+0.7}$    & $0.130_{-0.001}^{+0.002}$ & $0.0315_{-0.0008}^{+0.0008}$ & $0.0415_{-0.0002}^{+0.0003}$ & $0.04694_{-0.00008}^{+0.00008}$     \\ \hline
$\chi^2$/dof                                    & 950/956                   & 907/955                & 908/955                   & 916/952           & 940/954                   & 933/954           \\
$\tilde{\chi}^2$                                & 0.9936                    & 0.9502                 & 0.9504                    & 0.9618            & 0.9854                     & 0.9775             \\
$F_{2-10}$ ($10^{-9}$~ergs cm$^{-2}$ s$^{-1}$)  & $9.656_{-0.01}^{+0.01}$   & $9.652_{-0.010}^{+0.010}$  & $9.804_{-0.010}^{+0.010}$   & $9.824_{-0.010}^{+0.010}$  & $9.832_{-0.010}^{+0.010}$  & $9.807_{-0.010}^{+0.010}$        \\ \hline \hline
\end{tabular}
\leftline{$^f$ indicates parameters that were fixed during analysis.}
\leftline{`norm' refers to the normalization of the primary model component.}
\end{table*}
}

\section{Concluding Remarks}
\label{sec:conclusion}
We analyzed the \source~data obtained from MAXI/GSC in the form of one-day binned lightcurves, two NICER/XTI observations taken during the rising phase of the outburst, and one NuSTAR/FPMB observation taken near the peak of the outburst. Spectral analysis was performed on the NICER observations in the $0.3-10$ keV range in order to understand the parameters of the system at the beginning of the outburst. We then examined the NuSTAR observation in the $3-78$ keV range to more closely probe the accretion process and to measure parameters such as spin, inner disc radius, and inclination angle. 

The long term lightcurve analysis revealed that, in total, \source\ devoured roughly 90\% of the mass of Mars and released $\sim 5.2\times10^{44}$ ergs of energy in $2-20$ keV energy band. We also find that lightcurves in various energy bands correlate with each other, where softer photons are delayed compared to their harder counterparts. The maximum delay was observed between the $2-4$ and $10-20$ keV photons with a mean lag of $\tilde{\tau}=(8.4 \pm 1.9)$ days. Production of such a delay is likely related to the viscous delay of matter spiraling through the accretion disc during the outburst. This is supported by the presence of a strong multi-color disc component in the NICER spectra during the rising phase, with an increase normalization between the first and second observations.

Various models were used to understand the key parameters involved in the accretion process. In resolving the spectrum of \source~during the primary NuSTAR observation we revealed a sudden decrease in hard emission, with power-law normalization decreasing from $3.3_{-0.5}^{+0.6}$ to $1.54_{-0.07}^{+0.07}$ ph keV$^{-1}$ cm$^{-2}$. Such a reduction could have have resulted from large amounts of ejecta being carried away from the corona in a relatively short amount of time, so-called {\it blobby} jets. These ejections can appear to move at superluminal speeds, and have been definitively reported on in several other outbursting black holes \citep[e.g.][]{Fender1998,Hjellming1995}. Turning to the time-averaged spectrum, we obtained a hydrogen column density of $4.63^{+0.28}_{-0.28}\times10^{-22}$ cm$^{-2}$ using the {\tt WILM} \citep{Wilms2000} abundance from the {\sc Tbabs*\{Diskbb+Powerlaw\}} model. Using the {\tt ANGR} \citep{Anders1989} abundance, we found a lower ($2.99^{+0.11}_{-0.11}\times10^{-22}$ cm$^{-2}$) column density for the same model. The disc temperature varied from 1.15 to 1.23 keV across all the models, with {\sc Powerlaw} yielding an inner temperature of $T_{in}=1.18^{+0.015}_{-0.010}$~keV. From the {\sc Nthcomp} model, we estimated a coronal electron temperature of $kT_e=132^{+476}_{-123}$~keV. The coronal temperature remains unconstrained from our time-averaged spectral analysis. However, when we examined the hardness-resolved spectra using the {\sc Cutoffpl} model, we found the cutoff temperatures were better constrained, having values $E_{\rm cut} = 94_{-34}^{+78}$ keV in region i and $E_{\rm cut} = 79_{-18}^{+28}$ keV in region ii. We found signatures of the Fe K$\alpha$ line around $(6.48 \pm 0.02)$ keV with an equivalent width of $53$ eV. Apart from the Fe K$\alpha$ line, we also observed an Ni emission line around $(8.0 \pm 0.3)$ keV with an equivalent width of $60$ eV. This Ni line has previously been observed in several BHXRBs and AGNs \citep{Corliss1981,Molendi2003,Fukazawa2016,Medvedev2018}. It is possible that the $8.06_{-0.28}^{+0.10}$ keV line is observed due to the blue-shifted wing of the Fe K$\alpha$. However, if this were the case, we would also expect to observe a corresponding redshifted wing at roughly 6.4 - (8 - 6.4) = 4.8 keV \citep{Cui2000}. Given that we detect strong signatures of the emission line at 6.4 keV in the first place and the 8 keV Ni line has been previously identified, we believe that Ni emission line could be the more plausible explanation. Future generation satellites, such as \textit{Colibr\`i}, \citep{Heyl2019, Caiazzo2019} would be capable of resolving these lines more accurately and locating line emitting regions.

We applied phenomenological models, like {\sc Cutoffpl} to fit the NuSTAR data and estimate the luminosity of the accretion disc during that observation. Using $L_{2-10} = 4\pi D^2F_{2-10}$, assuming a distance of 10 kpc, we find $L_{2-10} = 1.09\times 10^{38}~\rm{erg~s^{-1}}$. Assuming a mass-to-radiation conversion efficiency of $\eta = 0.1$ based on our measured spin value \citep{Reynolds2021}, we find a mass accretion rate of $\dot{M} = L/\eta c^2 = 1.2\times 10^{18}~\rm{g~s^{-1}}$. Using the lower mass limit of \source~reported by \cite{Saha2022}, $\sim 5 M_{\odot}$ accounting for spin, we estimate $L/L_{\rm Edd} \sim 0.17$ during the observation. Combining this mass with the MAXI/GSC light curves, we find a maximum accretion rate of $L/L_{\rm Edd} \sim 0.28$ during the peak of the outburst.

Three models from the {\sc Relxill} family were selected for analysis with the NuSTAR data in order to estimate the more complex properties of \source, namely {\sc RelxillCp}, {\sc RelxillLp}, and {\sc RelxillLpCp}. Since {\sc RelxillCp} assumes that the irradiation of the disc behaves like a broken power-law, this model estimates a steep inner emissivity index of $q_1 = 8.61_{-0.41}^{+0.37}$ that becomes much flatter ($q_2 = 0.35_{-0.11}^{+0.10}$) at the break radius $R_{\rm br} = 14.3_{-0.6}^{+0.6}~r_g$. Otherwise, the three models estimate spin parameters of $0.65_{-0.06}^{+0.04}$, $0.70_{-0.31}^{+0.15}$, and $0.63_{-0.07}^{+0.07}$; inner disc radii of $2.2_{-0.6}^{+0.4}$~$R_{\mathrm{ISCO}}$,  $2.5_{-0.5}^{+0.5}~R_{\mathrm{ISCO}}$, and $1.7_{-0.5}^{+1.4}~R_{\mathrm{ISCO}}$; and inclination angles of $69.8_{-0.5}^{+0.5}$ degrees, $58_{-2}^{+2}$ degrees, and $37_{-3}^{+3}$ degrees, respectively. 

\section*{Acknowledgements}
 We acknowledge the anonymous reviewer for the helpful comments and suggestions which improved the paper. SH, AC, SSH and JH are supported by the Canadian Space Agency (CSA) and the Natural Sciences and Engineering Research Council of Canada (NSERC) through the Discovery Grants and the Canada Research Chairs programs. AJ acknowledge the support of the grant from the Ministry of Science and Technology of Taiwan with the grand number MOST 110-2811-M-007-500 and MOST 111-2811-M-007-002. This research made use of the {\it NuSTAR} Data Analysis Software ({\tt NuSTARDAS}) jointly developed by the ASI Space Science Data Center (ASSDC, Italy) and the California Institute of Technology (Caltech, USA).

\section*{Data Availability}
We used publicly available archival data of MAXI, NICER and NuSTAR observatories for this work. All the models used in this work, are publicly available. Appropriate links are provided in the text.



\bibliographystyle{mnras}
\bibliography{bhxrb} 

\begin{thebibliography}{}
\makeatletter
\relax
\def\mn@urlcharsother{\let\do\@makeother \do\$\do\&\do\#\do\^\do\_\do\%\do\~}
\def\mn@doi{\begingroup\mn@urlcharsother \@ifnextchar [ {\mn@doi@}
  {\mn@doi@[]}}
\def\mn@doi@[#1]#2{\def\@tempa{#1}\ifx\@tempa\@empty \href
  {http://dx.doi.org/#2} {doi:#2}\else \href {http://dx.doi.org/#2} {#1}\fi
  \endgroup}
\def\mn@eprint#1#2{\mn@eprint@#1:#2::\@nil}
\def\mn@eprint@arXiv#1{\href {http://arxiv.org/abs/#1} {{\tt arXiv:#1}}}
\def\mn@eprint@dblp#1{\href {http://dblp.uni-trier.de/rec/bibtex/#1.xml}
  {dblp:#1}}
\def\mn@eprint@#1:#2:#3:#4\@nil{\def\@tempa {#1}\def\@tempb {#2}\def\@tempc
  {#3}\ifx \@tempc \@empty \let \@tempc \@tempb \let \@tempb \@tempa \fi \ifx
  \@tempb \@empty \def\@tempb {arXiv}\fi \@ifundefined
  {mn@eprint@\@tempb}{\@tempb:\@tempc}{\expandafter \expandafter \csname
  mn@eprint@\@tempb\endcsname \expandafter{\@tempc}}}

\bibitem[\protect\citeauthoryear{{Alabarta} et~al.,}{{Alabarta}
  et~al.}{2021}]{Alberta2021}
{Alabarta} K.,  et~al., 2021, \mn@doi [\mnras] {10.1093/mnras/stab2241}, \href
  {https://ui.adsabs.harvard.edu/abs/2021MNRAS.507.5507A} {507, 5507}

\bibitem[\protect\citeauthoryear{{Alexander}}{{Alexander}}{2014}]{Tal2014}
{Alexander} T.,  2014, {ZDCF: Z-Transformed Discrete Correlation Function},
  Astrophysics Source Code Library, record ascl:1404.002 (\mn@eprint {ascl}
  {1404.002})

\bibitem[\protect\citeauthoryear{{Anders} \& {Grevesse}}{{Anders} \&
  {Grevesse}}{1989}]{Anders1989}
{Anders} E.,  {Grevesse} N.,  1989, \mn@doi [\gca]
  {10.1016/0016-7037(89)90286-X}, \href
  {https://ui.adsabs.harvard.edu/abs/1989GeCoA..53..197A} {53, 197}

\bibitem[\protect\citeauthoryear{{Arnaud}}{{Arnaud}}{1996}]{Arnaud1996}
{Arnaud} K.~A.,  1996, in {Jacoby} G.~H.,  {Barnes} J.,  eds,  Astronomical
  Society of the Pacific Conference Series Vol. 101, Astronomical Data Analysis
  Software and Systems V. p.~17

\bibitem[\protect\citeauthoryear{{Barthelmy} et~al.,}{{Barthelmy}
  et~al.}{2019}]{Barthelmy2019a}
{Barthelmy} S.~D.,  et~al., 2019, The Astronomer's Telegram, \href
  {https://ui.adsabs.harvard.edu/abs/2019ATel12436....1B} {12436, 1}

\bibitem[\protect\citeauthoryear{{Bhuvana}, {Radhika}, {Agrawal}, {Mandal}  \&
  {Nandi}}{{Bhuvana} et~al.}{2021}]{Bhuvana2021}
{Bhuvana} G.~R.,  {Radhika} D.,  {Agrawal} V.~K.,  {Mandal} S.,   {Nandi} A.,
  2021, \mn@doi [\mnras] {10.1093/mnras/staa4012}, \href
  {https://ui.adsabs.harvard.edu/abs/2021MNRAS.501.5457B} {501, 5457}

\bibitem[\protect\citeauthoryear{{Bright}, {Fender}, {Woudt}  \&
  {Miller-Jones}}{{Bright} et~al.}{2019}]{Bright2019}
{Bright} J.,  {Fender} R.,  {Woudt} P.,   {Miller-Jones} J.,  2019, The
  Astronomer's Telegram, \href
  {https://ui.adsabs.harvard.edu/abs/2019ATel12522....1B} {12522, 1}

\bibitem[\protect\citeauthoryear{Caiazzo et~al.,}{Caiazzo
  et~al.}{2019}]{Caiazzo2019}
Caiazzo I.,  et~al., 2019, {Unveiling the secrets of black holes and neutron
  stars with high-throughput, high-energy resolution X-ray spectroscopy},
  \mn@doi{10.5281/zenodo.3824441}, \url
  {https://doi.org/10.5281/zenodo.3824441}

\bibitem[\protect\citeauthoryear{{Casares} \& {Jonker}}{{Casares} \&
  {Jonker}}{2014}]{CJ2014}
{Casares} J.,  {Jonker} P.~G.,  2014, \mn@doi [\ssr]
  {10.1007/s11214-013-0030-6}, \href
  {https://ui.adsabs.harvard.edu/abs/2014SSRv..183..223C} {183, 223}

\bibitem[\protect\citeauthoryear{{Chakrabarti} \& {Titarchuk}}{{Chakrabarti} \&
  {Titarchuk}}{1995}]{CT95}
{Chakrabarti} S.,  {Titarchuk} L.~G.,  1995, \mn@doi [\apj] {10.1086/176610},
  \href {https://ui.adsabs.harvard.edu/abs/1995ApJ...455..623C} {455, 623}

\bibitem[\protect\citeauthoryear{{Chan}, {Psaltis}  \& {{\"O}zel}}{{Chan}
  et~al.}{2013}]{Chan2013}
{Chan} C.-k.,  {Psaltis} D.,   {{\"O}zel} F.,  2013, \mn@doi [\apj]
  {10.1088/0004-637X/777/1/13}, \href
  {https://ui.adsabs.harvard.edu/abs/2013ApJ...777...13C} {777, 13}

\bibitem[\protect\citeauthoryear{Chatterjee, Dutta, Nandi  \&
  Chakrabarti}{Chatterjee et~al.}{2020}]{Chatterjee2020}
Chatterjee A.,  Dutta B.~G.,  Nandi P.,   Chakrabarti S.~K.,  2020, \mn@doi
  [\mnras] {10.1093/mnras/staa2263}, 497, 4222

\bibitem[\protect\citeauthoryear{{Churazov}, {Gilfanov}  \&
  {Revnivtsev}}{{Churazov} et~al.}{2001}]{Churazov2001}
{Churazov} E.,  {Gilfanov} M.,   {Revnivtsev} M.,  2001, \mn@doi [\mnras]
  {10.1046/j.1365-8711.2001.04056.x}, \href
  {https://ui.adsabs.harvard.edu/abs/2001MNRAS.321..759C} {321, 759}

\bibitem[\protect\citeauthoryear{{Connors} et~al.,}{{Connors}
  et~al.}{2022}]{Connors2022}
{Connors} R. M.~T.,  et~al., 2022, \mn@doi [\apj] {10.3847/1538-4357/ac7ff2},
  \href {https://ui.adsabs.harvard.edu/abs/2022ApJ...935..118C} {935, 118}

\bibitem[\protect\citeauthoryear{Corliss \& Sugar}{Corliss \&
  Sugar}{1981}]{Corliss1981}
Corliss C.,  Sugar J.,  1981, \mn@doi [Journal of Physical and Chemical
  Reference Data] {10.1063/1.555638}, 10, 197

\bibitem[\protect\citeauthoryear{{Corral-Santana}, {Casares},
  {Mu{\~n}oz-Darias}, {Bauer}, {Mart{\'\i}nez-Pais}  \&
  {Russell}}{{Corral-Santana} et~al.}{2016}]{Corral2016}
{Corral-Santana} J.~M.,  {Casares} J.,  {Mu{\~n}oz-Darias} T.,  {Bauer} F.~E.,
  {Mart{\'\i}nez-Pais} I.~G.,   {Russell} D.~M.,  2016, \mn@doi [\aap]
  {10.1051/0004-6361/201527130}, \href
  {https://ui.adsabs.harvard.edu/abs/2016A&A...587A..61C} {587, A61}

\bibitem[\protect\citeauthoryear{{Cui}, {Chen}  \& {Zhang}}{{Cui}
  et~al.}{2000}]{Cui2000}
{Cui} W.,  {Chen} W.,   {Zhang} S.~N.,  2000, \mn@doi [\apj] {10.1086/308314},
  \href {https://ui.adsabs.harvard.edu/abs/2000ApJ...529..952C} {529, 952}

\bibitem[\protect\citeauthoryear{{Dauser}, {Garcia}, {Parker}, {Fabian}  \&
  {Wilms}}{{Dauser} et~al.}{2014}]{Dauser2014}
{Dauser} T.,  {Garcia} J.,  {Parker} M.~L.,  {Fabian} A.~C.,   {Wilms} J.,
  2014, \mn@doi [\mnras] {10.1093/mnrasl/slu125}, \href
  {https://ui.adsabs.harvard.edu/abs/2014MNRAS.444L.100D} {444, L100}

\bibitem[\protect\citeauthoryear{{Dauser}, {Garc{\'\i}a}, {Walton}, {Eikmann},
  {Kallman}, {McClintock}  \& {Wilms}}{{Dauser} et~al.}{2016}]{Dauser2016}
{Dauser} T.,  {Garc{\'\i}a} J.,  {Walton} D.~J.,  {Eikmann} W.,  {Kallman} T.,
  {McClintock} J.,   {Wilms} J.,  2016, \mn@doi [\aap]
  {10.1051/0004-6361/201628135}, \href
  {https://ui.adsabs.harvard.edu/abs/2016A&A...590A..76D} {590, A76}

\bibitem[\protect\citeauthoryear{{Done}, {Gierli{\'n}ski}  \& {Kubota}}{{Done}
  et~al.}{2007}]{Done2007}
{Done} C.,  {Gierli{\'n}ski} M.,   {Kubota} A.,  2007, \mn@doi [\aapr]
  {10.1007/s00159-007-0006-1}, \href
  {https://ui.adsabs.harvard.edu/abs/2007A&ARv..15....1D} {15, 1}

\bibitem[\protect\citeauthoryear{{Draghis}, {Miller}, {Zoghbi}, {Kammoun},
  {Reynolds}  \& {Tomsick}}{{Draghis} et~al.}{2021}]{Draghis2021}
{Draghis} P.~A.,  {Miller} J.~M.,  {Zoghbi} A.,  {Kammoun} E.~S.,  {Reynolds}
  M.~T.,   {Tomsick} J.~A.,  2021, \mn@doi [\apj] {10.3847/1538-4357/ac1270},
  \href {https://ui.adsabs.harvard.edu/abs/2021ApJ...920...88D} {920, 88}

\bibitem[\protect\citeauthoryear{{Draghis}, {Miller}, {Zoghbi}, {Reynolds},
  {Costantini}, {Gallo}  \& {Tomsick}}{{Draghis} et~al.}{2022}]{Draghis2022}
{Draghis} P.~A.,  {Miller} J.~M.,  {Zoghbi} A.,  {Reynolds} M.,  {Costantini}
  E.,  {Gallo} L.~C.,   {Tomsick} J.~A.,  2022, arXiv e-prints, \href
  {https://ui.adsabs.harvard.edu/abs/2022arXiv221002479D} {p. arXiv:2210.02479}

\bibitem[\protect\citeauthoryear{{Ducci} et~al.,}{{Ducci}
  et~al.}{2019}]{Ducci2019}
{Ducci} L.,  et~al., 2019, The Astronomer's Telegram, \href
  {https://ui.adsabs.harvard.edu/abs/2019ATel12502....1D} {12502, 1}

\bibitem[\protect\citeauthoryear{{Enoto} et~al.,}{{Enoto}
  et~al.}{2019}]{Enoto2019}
{Enoto} T.,  et~al., 2019, The Astronomer's Telegram, \href
  {https://ui.adsabs.harvard.edu/abs/2019ATel12455....1E} {12455, 1}

\bibitem[\protect\citeauthoryear{{Event Horizon Telescope Collaboration}
  et~al.,}{{Event Horizon Telescope Collaboration} et~al.}{2019}]{EHT2019}
{Event Horizon Telescope Collaboration} et~al., 2019, \mn@doi [\apjl]
  {10.3847/2041-8213/ab0ec7}, \href
  {https://ui.adsabs.harvard.edu/abs/2019ApJ...875L...1E} {875, L1}

\bibitem[\protect\citeauthoryear{{Event Horizon Telescope Collaboration}
  et~al.,}{{Event Horizon Telescope Collaboration} et~al.}{2022}]{EHT2022}
{Event Horizon Telescope Collaboration} et~al., 2022, \mn@doi [\apjl]
  {10.3847/2041-8213/ac6674}, \href
  {https://ui.adsabs.harvard.edu/abs/2022ApJ...930L..12E} {930, L12}

\bibitem[\protect\citeauthoryear{{Fabian}, {Rees}, {Stella}  \&
  {White}}{{Fabian} et~al.}{1989}]{Fabian1989}
{Fabian} A.~C.,  {Rees} M.~J.,  {Stella} L.,   {White} N.~E.,  1989, \mn@doi
  [\mnras] {10.1093/mnras/238.3.729}, \href
  {https://ui.adsabs.harvard.edu/abs/1989MNRAS.238..729F} {238, 729}

\bibitem[\protect\citeauthoryear{{Fabian}, {Lohfink}, {Kara}, {Parker},
  {Vasudevan}  \& {Reynolds}}{{Fabian} et~al.}{2015}]{Fabian2015}
{Fabian} A.~C.,  {Lohfink} A.,  {Kara} E.,  {Parker} M.~L.,  {Vasudevan} R.,
  {Reynolds} C.~S.,  2015, \mn@doi [\mnras] {10.1093/mnras/stv1218}, \href
  {https://ui.adsabs.harvard.edu/abs/2015MNRAS.451.4375F} {451, 4375}

\bibitem[\protect\citeauthoryear{Fender, Garrington, Mckay, Muxlow, Pooley,
  Spencer, Stirling  \& Waltman}{Fender et~al.}{1998}]{Fender1998}
Fender R.~P.,  Garrington S.~T.,  Mckay D.~J.,  Muxlow T. W.~B.,  Pooley G.~G.,
   Spencer R.~E.,  Stirling A.~M.,   Waltman E.~B.,  1998, Mon. Not. R. Astron.
  Soc, 000, 0

\bibitem[\protect\citeauthoryear{Foreman-Mackey}{Foreman-Mackey}{2016}]{corner}
Foreman-Mackey D.,  2016, \mn@doi [The Journal of Open Source Software]
  {10.21105/joss.00024}, 1, 24

\bibitem[\protect\citeauthoryear{{Fukazawa}, {Furui}, {Hayashi}, {Ohno},
  {Hiragi}  \& {Noda}}{{Fukazawa} et~al.}{2016}]{Fukazawa2016}
{Fukazawa} Y.,  {Furui} S.,  {Hayashi} K.,  {Ohno} M.,  {Hiragi} K.,   {Noda}
  H.,  2016, \mn@doi [\apj] {10.3847/0004-637X/821/1/15}, \href
  {https://ui.adsabs.harvard.edu/abs/2016ApJ...821...15F} {821, 15}

\bibitem[\protect\citeauthoryear{{Garain}, {Ghosh}  \& {Chakrabarti}}{{Garain}
  et~al.}{2012}]{Ghosh2012}
{Garain} S.~K.,  {Ghosh} H.,   {Chakrabarti} S.~K.,  2012, \mn@doi [\apj]
  {10.1088/0004-637X/758/2/114}, \href
  {https://ui.adsabs.harvard.edu/abs/2012ApJ...758..114G} {758, 114}

\bibitem[\protect\citeauthoryear{{Garain}, {Ghosh}  \& {Chakrabarti}}{{Garain}
  et~al.}{2014}]{Garain2014}
{Garain} S.~K.,  {Ghosh} H.,   {Chakrabarti} S.~K.,  2014, \mn@doi [\mnras]
  {10.1093/mnras/stt1969}, \href
  {https://ui.adsabs.harvard.edu/abs/2014MNRAS.437.1329G} {437, 1329}

\bibitem[\protect\citeauthoryear{{Garc{\'\i}a} \& {Kallman}}{{Garc{\'\i}a} \&
  {Kallman}}{2010}]{Garcia2010}
{Garc{\'\i}a} J.,  {Kallman} T.~R.,  2010, \mn@doi [\apj]
  {10.1088/0004-637X/718/2/695}, \href
  {https://ui.adsabs.harvard.edu/abs/2010ApJ...718..695G} {718, 695}

\bibitem[\protect\citeauthoryear{{Garc{\'\i}a}, {Dauser}, {Reynolds},
  {Kallman}, {McClintock}, {Wilms}  \& {Eikmann}}{{Garc{\'\i}a}
  et~al.}{2013}]{Garcia2013}
{Garc{\'\i}a} J.,  {Dauser} T.,  {Reynolds} C.~S.,  {Kallman} T.~R.,
  {McClintock} J.~E.,  {Wilms} J.,   {Eikmann} W.,  2013, \mn@doi [\apj]
  {10.1088/0004-637X/768/2/146}, \href
  {https://ui.adsabs.harvard.edu/abs/2013ApJ...768..146G} {768, 146}

\bibitem[\protect\citeauthoryear{{Garc{\'\i}a} et~al.,}{{Garc{\'\i}a}
  et~al.}{2014}]{Garcia2014}
{Garc{\'\i}a} J.,  et~al., 2014, \mn@doi [\apj] {10.1088/0004-637X/782/2/76},
  \href {https://ui.adsabs.harvard.edu/abs/2014ApJ...782...76G} {782, 76}

\bibitem[\protect\citeauthoryear{{Garc{\'\i}a} et~al.,}{{Garc{\'\i}a}
  et~al.}{2018}]{Garcia2018}
{Garc{\'\i}a} J.~A.,  et~al., 2018, \mn@doi [\apj] {10.3847/1538-4357/aad231},
  \href {https://ui.adsabs.harvard.edu/abs/2018ApJ...864...25G} {864, 25}

\bibitem[\protect\citeauthoryear{{Garc{\'\i}a} et~al.,}{{Garc{\'\i}a}
  et~al.}{2019}]{Garcia2019}
{Garc{\'\i}a} J.~A.,  et~al., 2019, \mn@doi [\apj] {10.3847/1538-4357/ab384f},
  \href {https://ui.adsabs.harvard.edu/abs/2019ApJ...885...48G} {885, 48}

\bibitem[\protect\citeauthoryear{{Gendreau}, {Arzoumanian}  \&
  {Okajima}}{{Gendreau} et~al.}{2012}]{Gendreau2012}
{Gendreau} K.~C.,  {Arzoumanian} Z.,   {Okajima} T.,  2012, in {Takahashi} T.,
  {Murray} S.~S.,   {den Herder} J.-W.~A.,  eds,  Society of Photo-Optical
  Instrumentation Engineers (SPIE) Conference Series Vol. 8443, Space
  Telescopes and Instrumentation 2012: Ultraviolet to Gamma Ray. p. 844313,
  \mn@doi{10.1117/12.926396}

\bibitem[\protect\citeauthoryear{{George} \& {Fabian}}{{George} \&
  {Fabian}}{1991}]{George1991}
{George} I.~M.,  {Fabian} A.~C.,  1991, \mn@doi [\mnras]
  {10.1093/mnras/249.2.352}, \href
  {https://ui.adsabs.harvard.edu/abs/1991MNRAS.249..352G} {249, 352}

\bibitem[\protect\citeauthoryear{{Gou} et~al.,}{{Gou} et~al.}{2009}]{Gou2009}
{Gou} L.,  et~al., 2009, \mn@doi [\apj] {10.1088/0004-637X/701/2/1076}, \href
  {https://ui.adsabs.harvard.edu/abs/2009ApJ...701.1076G} {701, 1076}

\bibitem[\protect\citeauthoryear{{Harrison} et~al.,}{{Harrison}
  et~al.}{2013}]{Harrison2013}
{Harrison} F.~A.,  et~al., 2013, \mn@doi [\apj] {10.1088/0004-637X/770/2/103},
  \href {https://ui.adsabs.harvard.edu/abs/2013ApJ...770..103H} {770, 103}

\bibitem[\protect\citeauthoryear{{Heyl} et~al.,}{{Heyl}
  et~al.}{2019}]{Heyl2019}
{Heyl} J.,  et~al., 2019, in Bulletin of the American Astronomical Society.
  p.~175

\bibitem[\protect\citeauthoryear{Hjellming \& Rupen}{Hjellming \&
  Rupen}{1995}]{Hjellming1995}
Hjellming R.~M.,  Rupen M.~P.,  1995, \mn@doi [Nature] {10.1038/375464a0}, 375,
  464

\bibitem[\protect\citeauthoryear{{Homan}, {Wijnands}, {van der Klis},
  {Belloni}, {van Paradijs}, {Klein-Wolt}, {Fender}  \& {M{\'e}ndez}}{{Homan}
  et~al.}{2001}]{Homan2001}
{Homan} J.,  {Wijnands} R.,  {van der Klis} M.,  {Belloni} T.,  {van Paradijs}
  J.,  {Klein-Wolt} M.,  {Fender} R.,   {M{\'e}ndez} M.,  2001, \mn@doi [\apjs]
  {10.1086/318954}, \href
  {https://ui.adsabs.harvard.edu/abs/2001ApJS..132..377H} {132, 377}

\bibitem[\protect\citeauthoryear{{Horne}}{{Horne}}{1985}]{Horne1985}
{Horne} K.,  1985, \mn@doi [\mnras] {10.1093/mnras/213.2.129}, \href
  {https://ui.adsabs.harvard.edu/abs/1985MNRAS.213..129H} {213, 129}

\bibitem[\protect\citeauthoryear{{Hu} et~al.,}{{Hu} et~al.}{2019}]{Hu2019}
{Hu} Y.~D.,  et~al., 2019, The Astronomer's Telegram, \href
  {https://ui.adsabs.harvard.edu/abs/2019ATel12443....1H} {12443, 1}

\bibitem[\protect\citeauthoryear{{Ingram}, {Done}  \& {Fragile}}{{Ingram}
  et~al.}{2009}]{Ingram2009}
{Ingram} A.,  {Done} C.,   {Fragile} P.~C.,  2009, \mn@doi [\mnras]
  {10.1111/j.1745-3933.2009.00693.x}, \href
  {https://ui.adsabs.harvard.edu/abs/2009MNRAS.397L.101I} {397, L101}

\bibitem[\protect\citeauthoryear{{Jana}, {Naik}, {Chatterjee}  \&
  {Jaisawal}}{{Jana} et~al.}{2021a}]{AJ2021d}
{Jana} A.,  {Naik} S.,  {Chatterjee} D.,   {Jaisawal} G.~K.,  2021a, \mn@doi
  [\mnras] {10.1093/mnras/stab2448}, \href
  {https://ui.adsabs.harvard.edu/abs/2021MNRAS.507.4779J} {507, 4779}

\bibitem[\protect\citeauthoryear{{Jana} et~al.,}{{Jana}
  et~al.}{2021b}]{AJ2021a}
{Jana} A.,  et~al., 2021b, \mn@doi [Research in Astronomy and Astrophysics]
  {-}, \href {-} {in Press., }

\bibitem[\protect\citeauthoryear{{Jana}, {Naik}, {Jaisawal}, {Chhotaray},
  {Kumari}  \& {Gupta}}{{Jana} et~al.}{2022}]{Jana2022}
{Jana} A.,  {Naik} S.,  {Jaisawal} G.~K.,  {Chhotaray} B.,  {Kumari} N.,
  {Gupta} S.,  2022, \mn@doi [\mnras] {10.1093/mnras/stac315}, \href
  {https://ui.adsabs.harvard.edu/abs/2022MNRAS.511.3922J} {511, 3922}

\bibitem[\protect\citeauthoryear{{Kreidberg}, {Bailyn}, {Farr}  \&
  {Kalogera}}{{Kreidberg} et~al.}{2012}]{Kreidberg2012}
{Kreidberg} L.,  {Bailyn} C.~D.,  {Farr} W.~M.,   {Kalogera} V.,  2012, \mn@doi
  [\apj] {10.1088/0004-637X/757/1/36}, \href
  {https://ui.adsabs.harvard.edu/abs/2012ApJ...757...36K} {757, 36}

\bibitem[\protect\citeauthoryear{{Kubota}, {Tanaka}, {Makishima}, {Ueda},
  {Dotani}, {Inoue}  \& {Yamaoka}}{{Kubota} et~al.}{1998}]{Kubota1998}
{Kubota} A.,  {Tanaka} Y.,  {Makishima} K.,  {Ueda} Y.,  {Dotani} T.,  {Inoue}
  H.,   {Yamaoka} K.,  1998, \mn@doi [\pasj] {10.1093/pasj/50.6.667}, \href
  {https://ui.adsabs.harvard.edu/abs/1998PASJ...50..667K} {50, 667}

\bibitem[\protect\citeauthoryear{{Li}, {Zimmerman}, {Narayan}  \&
  {McClintock}}{{Li} et~al.}{2005}]{Li2005}
{Li} L.-X.,  {Zimmerman} E.~R.,  {Narayan} R.,   {McClintock} J.~E.,  2005,
  \mn@doi [\apjs] {10.1086/428089}, \href
  {https://ui.adsabs.harvard.edu/abs/2005ApJS..157..335L} {157, 335}

\bibitem[\protect\citeauthoryear{{Luminet}}{{Luminet}}{1979}]{Luminet1979}
{Luminet} J.~P.,  1979, \aap, \href
  {https://ui.adsabs.harvard.edu/abs/1979A&A....75..228L} {75, 228}

\bibitem[\protect\citeauthoryear{{Magdziarz} \& {Zdziarski}}{{Magdziarz} \&
  {Zdziarski}}{1995}]{MZ1995}
{Magdziarz} P.,  {Zdziarski} A.~A.,  1995, \mn@doi [\mnras]
  {10.1093/mnras/273.3.837}, \href
  {https://ui.adsabs.harvard.edu/abs/1995MNRAS.273..837M} {273, 837}

\bibitem[\protect\citeauthoryear{{Mastroserio}, {Ingram}  \& {van der
  Klis}}{{Mastroserio} et~al.}{2019}]{Mastroserio2019}
{Mastroserio} G.,  {Ingram} A.,   {van der Klis} M.,  2019, \mn@doi [\mnras]
  {10.1093/mnras/stz1727}, \href
  {https://ui.adsabs.harvard.edu/abs/2019MNRAS.488..348M} {488, 348}

\bibitem[\protect\citeauthoryear{{Matsuoka} et~al.,}{{Matsuoka}
  et~al.}{2009}]{Matsuoka2009}
{Matsuoka} M.,  et~al., 2009, \mn@doi [\pasj] {10.1093/pasj/61.5.999}, \href
  {https://ui.adsabs.harvard.edu/abs/2009PASJ...61..999M} {61, 999}

\bibitem[\protect\citeauthoryear{{McClintock}, {Shafee}, {Narayan},
  {Remillard}, {Davis}  \& {Li}}{{McClintock} et~al.}{2006}]{McClintock2006}
{McClintock} J.~E.,  {Shafee} R.,  {Narayan} R.,  {Remillard} R.~A.,  {Davis}
  S.~W.,   {Li} L.-X.,  2006, \mn@doi [\apj] {10.1086/508457}, \href
  {https://ui.adsabs.harvard.edu/abs/2006ApJ...652..518M} {652, 518}

\bibitem[\protect\citeauthoryear{{Medvedev}, {Khabibullin}, {Sazonov},
  {Churazov}  \& {Tsygankov}}{{Medvedev} et~al.}{2018}]{Medvedev2018}
{Medvedev} P.~S.,  {Khabibullin} I.~I.,  {Sazonov} S.~Y.,  {Churazov} E.~M.,
  {Tsygankov} S.~S.,  2018, \mn@doi [Astronomy Letters]
  {10.1134/S1063773718060038}, \href
  {https://ui.adsabs.harvard.edu/abs/2018AstL...44..390M} {44, 390}

\bibitem[\protect\citeauthoryear{{Miller} et~al.,}{{Miller}
  et~al.}{2012}]{Miller2012}
{Miller} J.~M.,  et~al., 2012, \mn@doi [\apjl] {10.1088/2041-8205/759/1/L6},
  \href {https://ui.adsabs.harvard.edu/abs/2012ApJ...759L...6M} {759, L6}

\bibitem[\protect\citeauthoryear{{Miller} et~al.,}{{Miller}
  et~al.}{2018}]{Miller2018}
{Miller} J.~M.,  et~al., 2018, \mn@doi [\apjl] {10.3847/2041-8213/aacc61},
  \href {https://ui.adsabs.harvard.edu/abs/2018ApJ...860L..28M} {860, L28}

\bibitem[\protect\citeauthoryear{{Miniutti} \& {Fabian}}{{Miniutti} \&
  {Fabian}}{2004}]{MF2004}
{Miniutti} G.,  {Fabian} A.~C.,  2004, \mn@doi [\mnras]
  {10.1111/j.1365-2966.2004.07611.x}, \href
  {https://ui.adsabs.harvard.edu/abs/2004MNRAS.349.1435M} {349, 1435}

\bibitem[\protect\citeauthoryear{Molendi, Bianchi  \& Matt}{Molendi
  et~al.}{2003}]{Molendi2003}
Molendi S.,  Bianchi S.,   Matt G.,  2003, \mn@doi [\mnras]
  {10.1046/j.1365-8711.2003.06783.x}, 343, L1

\bibitem[\protect\citeauthoryear{{Mudambi}, {Rao}, {Gudennavar}, {Misra}  \&
  {Bubbly}}{{Mudambi} et~al.}{2020}]{Mudambi2020}
{Mudambi} S.~P.,  {Rao} A.,  {Gudennavar} S.~B.,  {Misra} R.,   {Bubbly} S.~G.,
   2020, \mn@doi [\mnras] {10.1093/mnras/staa2656}, \href
  {https://ui.adsabs.harvard.edu/abs/2020MNRAS.498.4404M} {498, 4404}

\bibitem[\protect\citeauthoryear{{Negoro} et~al.,}{{Negoro}
  et~al.}{2019}]{Negoro19}
{Negoro} H.,  et~al., 2019, The Astronomer's Telegram, \href
  {https://ui.adsabs.harvard.edu/abs/2019ATel13256....1N} {13256, 1}

\bibitem[\protect\citeauthoryear{{Novikov} \& {Thorne}}{{Novikov} \&
  {Thorne}}{1973}]{Novikov1973}
{Novikov} I.~D.,  {Thorne} K.~S.,  1973, in Black Holes (Les Astres Occlus). pp
  343--450

\bibitem[\protect\citeauthoryear{{Remillard} \& {McClintock}}{{Remillard} \&
  {McClintock}}{2006}]{RM06}
{Remillard} R.~A.,  {McClintock} J.~E.,  2006, \mn@doi [\araa]
  {10.1146/annurev.astro.44.051905.092532}, \href
  {https://ui.adsabs.harvard.edu/abs/2006ARA&A..44...49R} {44, 49}

\bibitem[\protect\citeauthoryear{{Reynolds}}{{Reynolds}}{2021}]{Reynolds2021}
{Reynolds} C.~S.,  2021, \mn@doi [\araa] {10.1146/annurev-astro-112420-035022},
  \href {https://ui.adsabs.harvard.edu/abs/2021ARA&A..59..117R} {59, 117}

\bibitem[\protect\citeauthoryear{{Saha}, {Mandal}  \& {Pal}}{{Saha}
  et~al.}{2022}]{Saha2022}
{Saha} D.,  {Mandal} M.,   {Pal} S.,  2022, arXiv e-prints, \href
  {https://ui.adsabs.harvard.edu/abs/2022arXiv221013748S} {p. arXiv:2210.13748}

\bibitem[\protect\citeauthoryear{{Shakura} \& {Sunyaev}}{{Shakura} \&
  {Sunyaev}}{1973}]{SS73}
{Shakura} N.~I.,  {Sunyaev} R.~A.,  1973, \aap, \href
  {https://ui.adsabs.harvard.edu/abs/1973A&A....24..337S} {500, 33}

\bibitem[\protect\citeauthoryear{{Shaposhnikov} \& {Titarchuk}}{{Shaposhnikov}
  \& {Titarchuk}}{2007}]{Shaposhnikov2007}
{Shaposhnikov} N.,  {Titarchuk} L.,  2007, \mn@doi [\apj] {10.1086/518110},
  \href {https://ui.adsabs.harvard.edu/abs/2007ApJ...663..445S} {663, 445}

\bibitem[\protect\citeauthoryear{{Smith}, {Heindl}  \& {Swank}}{{Smith}
  et~al.}{2002}]{Smith2002_accretion}
{Smith} D.~M.,  {Heindl} W.~A.,   {Swank} J.~H.,  2002, \mn@doi [\apj]
  {10.1086/339167}, \href
  {https://ui.adsabs.harvard.edu/abs/2002ApJ...569..362S} {569, 362}

\bibitem[\protect\citeauthoryear{{Sreehari}, {Ravishankar}, {Iyer}, {Agrawal},
  {Katoch}, {Mandal}  \& {Nandi}}{{Sreehari} et~al.}{2019}]{Sreehari2019}
{Sreehari} H.,  {Ravishankar} B.~T.,  {Iyer} N.,  {Agrawal} V.~K.,  {Katoch}
  T.~B.,  {Mandal} S.,   {Nandi} A.,  2019, \mn@doi [\mnras]
  {10.1093/mnras/stz1327}, \href
  {https://ui.adsabs.harvard.edu/abs/2019MNRAS.487..928S} {487, 928}

\bibitem[\protect\citeauthoryear{Sreehari, Nandi, Das, Agrawal, Mandal,
  Ramadevi  \& Katoch}{Sreehari et~al.}{2020}]{Sree2020}
Sreehari H.,  Nandi A.,  Das S.,  Agrawal V.~K.,  Mandal S.,  Ramadevi M.~C.,
  Katoch T.,  2020, \mn@doi [\mnras] {10.1093/mnras/staa3135}, 499, 5891

\bibitem[\protect\citeauthoryear{{Steiner}, {Narayan}, {McClintock}  \&
  {Ebisawa}}{{Steiner} et~al.}{2009}]{Steiner2009}
{Steiner} J.~F.,  {Narayan} R.,  {McClintock} J.~E.,   {Ebisawa} K.,  2009,
  \mn@doi [\pasp] {10.1086/648535}, \href
  {https://ui.adsabs.harvard.edu/abs/2009PASP..121.1279S} {121, 1279}

\bibitem[\protect\citeauthoryear{{Steiner}, {McClintock}, {Orosz}, {Remillard},
  {Bailyn}, {Kolehmainen}  \& {Straub}}{{Steiner} et~al.}{2014}]{Steiner2014}
{Steiner} J.~F.,  {McClintock} J.~E.,  {Orosz} J.~A.,  {Remillard} R.~A.,
  {Bailyn} C.~D.,  {Kolehmainen} M.,   {Straub} O.,  2014, \mn@doi [\apjl]
  {10.1088/2041-8205/793/2/L29}, \href
  {https://ui.adsabs.harvard.edu/abs/2014ApJ...793L..29S} {793, L29}

\bibitem[\protect\citeauthoryear{{Sunyaev} \& {Titarchuk}}{{Sunyaev} \&
  {Titarchuk}}{1980}]{ST80}
{Sunyaev} R.~A.,  {Titarchuk} L.~G.,  1980, \aap, \href
  {https://ui.adsabs.harvard.edu/abs/1980A&A....86..121S} {500, 167}

\bibitem[\protect\citeauthoryear{{Sunyaev} \& {Titarchuk}}{{Sunyaev} \&
  {Titarchuk}}{1985}]{ST85}
{Sunyaev} R.~A.,  {Titarchuk} L.~G.,  1985, \aap, \href
  {https://ui.adsabs.harvard.edu/abs/1985A&A...143..374S} {143, 374}

\bibitem[\protect\citeauthoryear{{Taylor}, {Uttley}  \& {McHardy}}{{Taylor}
  et~al.}{2003}]{Taylor2003}
{Taylor} R.~D.,  {Uttley} P.,   {McHardy} I.~M.,  2003, \mn@doi [\mnras]
  {10.1046/j.1365-8711.2003.06742.x}, \href
  {https://ui.adsabs.harvard.edu/abs/2003MNRAS.342L..31T} {342, L31}

\bibitem[\protect\citeauthoryear{{Tetarenko}, {Sivakoff}, {Heinke}  \&
  {Gladstone}}{{Tetarenko} et~al.}{2016}]{Tetarenko2016}
{Tetarenko} B.~E.,  {Sivakoff} G.~R.,  {Heinke} C.~O.,   {Gladstone} J.~C.,
  2016, \mn@doi [\apjs] {10.3847/0067-0049/222/2/15}, \href
  {https://ui.adsabs.harvard.edu/abs/2016ApJS..222...15T} {222, 15}

\bibitem[\protect\citeauthoryear{{Tomsick} et~al.,}{{Tomsick}
  et~al.}{2018}]{Tomsick2018}
{Tomsick} J.~A.,  et~al., 2018, \mn@doi [\apj] {10.3847/1538-4357/aaaab1},
  \href {https://ui.adsabs.harvard.edu/abs/2018ApJ...855....3T} {855, 3}

\bibitem[\protect\citeauthoryear{{Torres}, {Casares}, {Jim{\'e}nez-Ibarra},
  {Mu{\~n}oz-Darias}, {Armas Padilla}, {Jonker}  \& {Heida}}{{Torres}
  et~al.}{2019}]{Torres2019}
{Torres} M.~A.~P.,  {Casares} J.,  {Jim{\'e}nez-Ibarra} F.,  {Mu{\~n}oz-Darias}
  T.,  {Armas Padilla} M.,  {Jonker} P.~G.,   {Heida} M.,  2019, \mn@doi
  [\apjl] {10.3847/2041-8213/ab39df}, \href
  {https://ui.adsabs.harvard.edu/abs/2019ApJ...882L..21T} {882, L21}

\bibitem[\protect\citeauthoryear{{Uttley et al.,}}{{Uttley et
  al.,}}{2020}]{Uttley2020}
{Uttley et al.,} P.,  2020, in Society of Photo-Optical Instrumentation
  Engineers (SPIE) Conference Series. p. 114441E, \mn@doi{10.1117/12.2562523}

\bibitem[\protect\citeauthoryear{{Verner}, {Ferland}, {Korista}  \&
  {Yakovlev}}{{Verner} et~al.}{1996}]{Verner1996}
{Verner} D.~A.,  {Ferland} G.~J.,  {Korista} K.~T.,   {Yakovlev} D.~G.,  1996,
  \mn@doi [\apj] {10.1086/177435}, \href
  {https://ui.adsabs.harvard.edu/abs/1996ApJ...465..487V} {465, 487}

\bibitem[\protect\citeauthoryear{{Viergutz}}{{Viergutz}}{1993}]{Viergutz1993}
{Viergutz} S.~U.,  1993, \aap, \href
  {https://ui.adsabs.harvard.edu/abs/1993A&A...272..355V} {272, 355}

\bibitem[\protect\citeauthoryear{{Wilms}, {Allen}  \& {McCray}}{{Wilms}
  et~al.}{2000}]{Wilms2000}
{Wilms} J.,  {Allen} A.,   {McCray} R.,  2000, \mn@doi [\apj] {10.1086/317016},
  \href {https://ui.adsabs.harvard.edu/abs/2000ApJ...542..914W} {542, 914}

\bibitem[\protect\citeauthoryear{{Xu}, {Harrison}, {Tomsick}, {Walton},
  {Barret}, {Garc{\'\i}a}, {Hare}  \& {Parker}}{{Xu} et~al.}{2020}]{Xu2020}
{Xu} Y.,  {Harrison} F.~A.,  {Tomsick} J.~A.,  {Walton} D.~J.,  {Barret} D.,
  {Garc{\'\i}a} J.~A.,  {Hare} J.,   {Parker} M.~L.,  2020, \mn@doi [\apj]
  {10.3847/1538-4357/ab7dc0}, \href
  {https://ui.adsabs.harvard.edu/abs/2020ApJ...893...30X} {893, 30}

\bibitem[\protect\citeauthoryear{Zdziarski \& Marco}{Zdziarski \&
  Marco}{2020}]{Zdziarski2020}
Zdziarski A.~A.,  Marco B.~D.,  2020, \mn@doi [The Astrophysical Journal]
  {10.3847/2041-8213/ab9899}, 896, L36

\bibitem[\protect\citeauthoryear{{Zdziarski}, {Johnson}  \&
  {Magdziarz}}{{Zdziarski} et~al.}{1996}]{Z96}
{Zdziarski} A.~A.,  {Johnson} W.~N.,   {Magdziarz} P.,  1996, \mn@doi [\mnras]
  {10.1093/mnras/283.1.193}, \href
  {https://ui.adsabs.harvard.edu/abs/1996MNRAS.283..193Z} {283, 193}

\bibitem[\protect\citeauthoryear{{Zhang}, {Cui}  \& {Chen}}{{Zhang}
  et~al.}{1997}]{Zhang1997}
{Zhang} S.~N.,  {Cui} W.,   {Chen} W.,  1997, \mn@doi [\apjl] {10.1086/310705},
  \href {https://ui.adsabs.harvard.edu/abs/1997ApJ...482L.155Z} {482, L155}

\bibitem[\protect\citeauthoryear{{Zhao} et~al.,}{{Zhao}
  et~al.}{2021}]{Zhao2021}
{Zhao} X.,  et~al., 2021, \mn@doi [\apj] {10.3847/1538-4357/ac07a9}, \href
  {https://ui.adsabs.harvard.edu/abs/2021ApJ...916..108Z} {916, 108}

\bibitem[\protect\citeauthoryear{{{\.Z}ycki}, {Done}  \& {Smith}}{{{\.Z}ycki}
  et~al.}{1999}]{Zycki1999}
{{\.Z}ycki} P.~T.,  {Done} C.,   {Smith} D.~A.,  1999, \mn@doi [\mnras]
  {10.1046/j.1365-8711.1999.02885.x}, \href
  {https://ui.adsabs.harvard.edu/abs/1999MNRAS.309..561Z} {309, 561}

\bibitem[\protect\citeauthoryear{den Hartog, Uttley, Willingale, Hoevers, den
  Herder  \& Wise}{den Hartog et~al.}{2020}]{Hartog2020}
den Hartog R.~H.,  Uttley P.,  Willingale R.,  Hoevers H.,  den Herder J.-W.,
  Wise M.,  2020, in den Herder J.-W.~A.,  Nakazawa K.,   Nikzad S.,  eds,
  Space Telescopes and Instrumentation 2020: Ultraviolet to Gamma Ray. {SPIE},
  \mn@doi{10.1117/12.2562831}, \url {https://doi.org/10.1117\%2F12.2562831}

\makeatother
\end{thebibliography}




\appendix


\bsp	
\label{lastpage}
\end{document}